# Supernova signals of light dark matter

William DeRocco,[1] Peter W. Graham,[1] Daniel Kasen,[2,3] Gustavo Marques-Tavares,[1,4] and Surjeet Rajendran[2]

[1]*Stanford Institute for Theoretical Physics, Stanford University, Stanford, California 94305, USA*
[2]*Berkeley Center for Theoretical Physics, Department of Physics, University of California, Berkeley, California 94720, USA*
[3]*Lawrence Berkeley National Laboratory, Berkeley, California 94720, USA*
[4]*Maryland Center for Fundamental Physics, Department of Physics, University of Maryland, College Park, Maryland 20742, USA*

Dark matter direct detection experiments have poor sensitivity to a galactic population of dark matter with mass below the GeV scale. However, such dark matter can be produced copiously in supernovae. Since this thermally produced population is much hotter than the galactic dark matter, it can be observed with direct detection experiments. In this paper, we focus on a dark sector with fermion dark matter and a heavy dark photon as a specific example. We first extend existing supernova cooling constraints on this model to the regime of strong coupling where the dark matter becomes diffusively trapped in the supernova. Then, using the fact that even outside these cooling constraints the diffuse galactic flux of these dark sector particles can still be large, we show that this flux is detectable in direct detection experiments such as current and next-generation liquid xenon detectors. As a result, due to supernova production, light dark matter has the potential to be discovered over many orders of magnitude of mass and coupling.

DOI: 10.1103/PhysRevD.100.075018

## I. INTRODUCTION

The particle nature of dark matter (DM) remains one of the largest outstanding puzzles in physics. Despite the overwhelming evidence for the existence of dark matter from its gravitational imprints on cosmological and astrophysical scales, there have as of yet been no observations of any nongravitational interactions [1]. The fact that our current measurements leave an enormous range of possibilities for its mass and interactions with the Standard Model (SM) has motivated a very rich experimental program exploring a wide range of dark matter models.

A large ongoing experimental effort is searching for dark matter candidates with masses in the GeV–TeV range (weakly-interacting massive particles, or "WIMPs"), largely motivated by the WIMP miracle (see, e.g., [2]). These experiments have achieved incredible sensitivity to dark matter by searching for very small energy depositions from dark matter scattering with nuclei in extremely clean environments. Despite their great progress, such experiments quickly lose sensitivity to lighter dark matter candidates because the kinetic energy of such candidates is too small to lead to observable signatures in the detectors. This is true under the assumption that we can only detect dark matter particles which are gravitationally bound to our galaxy, which implies a maximum velocity for the dark matter flux.

The existence of astrophysical sources where dark matter could be produced with larger velocities and at a non-negligible flux allows one to significantly extend the reach of these experiments to models of sub-GeV dark matter.[1] Core-collapse supernovae (SN) can reach core temperatures in excess of 30 MeV for $\mathcal{O}(10)$ seconds, allowing them to produce vast thermal fluxes of particles with masses $\lesssim \mathcal{O}(100)$ MeV at relativistic speeds. This makes them an ideal astrophysical source for sub-GeV dark matter. Supernovae have already been used extensively to constrain a plethora of models of new physics. In almost all cases, the criterion applied in order to place a bound is the so-called *cooling criterion*, which states that if any new particle were able to transport energy out of the proto-neutron star [7,8] formed by the SN more quickly than the neutrinos, the cooling timescale of the core would be less than the ten-second timescale observed in SN1987a. This is equivalent to the statement that any new particle that transports greater than $3 \times 10^{52}$ erg/s through the radius at which the neutrinos are no longer diffusively trapped is incompatible with observations [9].

---

[1]Another recent idea has been to use stars and cosmic rays to accelerate a fraction of the galactic DM to higher energies, also enhancing the sensitivity to some models of dark matter [3–6].

There are two distinct regimes relevant for these cooling bounds. At lower couplings one considers bulk emission from the entire protoneutron star volume, which results in a lower bound on the couplings below which the luminosity is less than $3 \times 10^{52}$ erg/s due to insufficient production of the new particles. There is a separate regime at large couplings, in which the coupling is large enough that the new particles are diffusively trapped inside the core and the emission effectively occurs from a radius at which the densities and temperatures are low enough to allow the new particles to escape freely. This is the regime in which the upper bound on the couplings constrained by cooling can be derived. The trapped regime is reasonably well understood analytically for particles that are singly emitted or absorbed (e.g., axions [10] and dark photons [11,12]). However, for particles that can only be pair produced, a detailed understanding of different processes (e.g., production versus scattering) is required, making analytic estimates more challenging. For examples of existing cooling constraints on pair-produced particles, see, e.g., Refs. [13–21].

## II. SUMMARY

In this paper, we move beyond these cooling arguments by considering the direct detection of a hot population of supernova-produced dark matter. Even in parameter space outside the cooling bound, a supernova can still produce a vast flux of light dark matter particles. This flux is also hot (semirelativistic), which allows for the possibility of its detection in current and next generation WIMP experiments. Using a Monte Carlo Boltzmann particle transport simulation, we are able to compute the DM flux in the trapped regime, which allows us to estimate the reach of direct detection experiments that were originally expected to only have sensitivity to dark matter with masses above $\sim$GeV. Furthermore, our results improve upon and extend previous cooling constraints in the trapped regime which relied on a number of approximations in order to produce an analytic estimate of the flux.

For concreteness we focus on a simple model of dark matter, in which it is a Dirac fermion which interacts with the SM via the four-fermion operator

$$\frac{e\epsilon g_d}{\Lambda^2} \bar{\chi}\gamma_\mu \chi J^\mu_{\rm em}, \qquad (1)$$

where $\chi$ is the DM field and $J^\mu_{\rm em}$ is the electromagnetic current of the SM. Such an interaction can be generated if, for example, DM is charged under a dark gauge boson with mass $m_{A'} = \Lambda$ and DM-DM coupling $g_d$ that kinetically mixes with the SM with mixing parameter $\epsilon$.[2] While this model is a good first test case, it should be noted that the main point of this paper (that SN can produce a flux of light dark matter that is observable in direct detection experiments) also applies to a much wider variety of models. We leave those for future work.

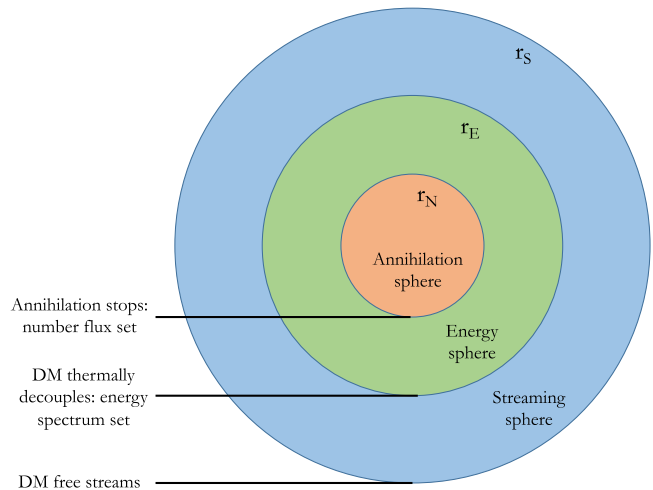

FIG. 1. The characteristic spheres of the protoneutron star. Outside the annihilation sphere, number-changing processes for the DM (pair production and bremsstrahlung) freeze out, setting the number flux. Outside the energy sphere, the DM thermally decouples and its spectrum is set. Outside the streaming sphere, the DM is no longer diffusively trapped by proton scattering and free-streams out of the star.

The high temperatures reached in core-collapse SN allow them to produce large abundances of the sub-GeV dark sector fermions considered in this paper. In the regions of parameter space we are interested in, these fermions have a sufficiently strong coupling to the Standard Model that they become diffusively trapped near the protoneutron star that forms from the SN core. The diffusive trapping is primarily due to scatterings off of the free protons generated by the dissociation of nuclei in the SN shock. These scattering interactions are inefficient at changing the dark fermion's energy because of the large mass ratio between the DM and the nuclei. The dark fermions also scatter off of electrons and positrons, allowing thermal exchange with the SM bath. At a certain radius (which we call the *electron sphere* or *energy sphere*) the density of electrons and positrons drops to the point where a dark fermion no longer remains in thermal contact with the SM; see Fig. 1. Additionally, there is a *streaming sphere* at which the density of protons drops low enough that dark fermions are no longer diffusively trapped and begin to free-stream out of the star. Any given dark fermion produced in the SN will therefore diffuse through a proton-rich overburden until it either reaches the streaming sphere and escapes or encounters an antiparticle and annihilates.

---

[2]If the new particle accounts for all of dark matter, there are stringent bounds from the cosmic microwave background (CMB) from DM late annihilation that place strong constraints on this model [22]. Those can be evaded by considering either asymmetric DM or by introducing a small mass splitting, making it a pseudo-Dirac fermion (see, e.g., [23,24]). Both of those scenarios would not affect any of our conclusions (as long as the mass splitting is small compared to the temperature of the SN), and so for simplicity, we choose to focus on the simple Dirac fermion model in this paper.

The dark fermions that do eventually escape are produced with a distribution of semirelativistic velocities. This results in a *time-spreading* effect during their propagation to Earth. The difference in arrival time of the high-momentum and low-momentum ends of the spectrum is of order the dark fermion travel time between Earth and the SN; hence the dark fermions produced by a single SN arrive on Earth over a timescale of $10^5$ years for an average galactic SN. This is in direct contrast to the neutrinos, which are all produced highly relativistically and therefore arrive as a single pulse over a timescale of ten seconds.

Given the typical rates and distances of galactic supernovae, in addition to the inherent signal spread, the dark fermion fluxes from various SN should overlap in time, producing a diffuse galactic SN flux of dark fermions. This signal is reminiscent of the diffuse flux of SN neutrinos, with the distinction that diffuse neutrinos would arise from the sum of a much greater number of extragalactic sources each of short duration, since the time between galactic SN is much larger than the duration of the neutrino flux.

If a diffuse dark matter SN flux is continuously passing through Earth, we must consider ways to detect it with Earth-based experiments. We find that the diffuse flux of DM is detectable in existing and next-generation liquid xenon (LXe) WIMP detectors. Interestingly, though the idea is similar to the direct detection of the diffuse SN neutrino flux, it is not the large neutrino detectors which are best to search for this flux but rather the WIMP detectors due to their low-energy thresholds. Though the WIMP detectors were designed to hunt for dark matter on the GeV scale, we show that they are sensitive to recoils by sub-GeV dark sector fermions over a wide range of masses and couplings above even the newly computed trapped regime cooling bound. As this idea probes extremely weak couplings, it is complementary to most of the experimental proposals searching for sub-GeV dark matter through direct detection or in accelerators (see [25,26] and references therein for details of some of the other proposals for detecting sub-GeV DM).

Existing LXe experiments such as Xenon 1T [27] are already sensitive to the diffuse galactic SN flux of dark sector fermions, and future experiments such as Xenon nT [28], PandaX-4T [29], LUX-Zeplin [30], and, on a longer time scale, DARWIN [31] will cover an even larger region of parameter space.

In Secs. III and IV, we describe an analytic treatment of the required computation and explain the details of the Monte Carlo Boltzmann transport simulation. We discuss our computation of new cooling bounds in Sec. V and direct detection by LXe detectors in Sec. VI. Our results are presented in Sec. VII.

### III. ANALYTIC APPROXIMATION

While the final results used in this paper were computed using a numerical Monte Carlo simulation of particle transport within the supernova and cooling protoneutron star, we first provide a simple physical picture of the expected behavior of the DM flux in the diffusive regime. We then apply the intuitive description to demonstrate how to make rough analytical estimates of the spectrum of the SN-produced dark fermion flux.

Our basic premise is to generalize the idea of a "neutrino sphere" to the case of dark fermions. The term "neutrino sphere" is typically used to describe the radius at which the density of nucleons has dropped such that the neutrinos are no longer diffusively trapped. It is common to approximate the SN neutrino flux as simply the emission of a blackbody sphere with radius and temperature given at the neutrino sphere. This approximation is reasonable for electron neutrinos and antineutrinos since beta processes are the dominant interaction maintaining thermal equilibrium and causing diffusive trapping of these neutrinos. Hence, when this interaction ceases to become efficient due to falling nucleon density, the neutrinos will free-stream from the same radius at which their temperature has been set [32].

However, this is not the case for mu and tau neutrinos [33]. As shown in Ref. [32], second and third generation neutrinos are kept in thermal equilibrium by nuclear bremsstrahlung ($NN \leftrightarrow NN\nu\nu$) but are kept diffusively trapped by nuclear scatterings ($\nu N \rightarrow \nu N$). These interactions freeze out at different radii; hence their flux must be computed with a combination of a blackbody emission plus a transmission calculation through a scattering atmosphere [32]. Note that since bremsstrahlung keeps the neutrinos in thermal equilibrium, its freeze-out sets both the number flux and temperature of the outgoing neutrinos.

In the case of the dark fermions, the interactions that set the number flux, the energy spectrum (temperature), and the free-streaming radius are all different, so, in contrast to the mu and tau neutrinos, there are now *three* distinct radii, one for each of these interactions.[3] We therefore break the protoneutron star into three radii we have termed *characteristic spheres* at which different interactions freeze

---

[3]The main reason for this difference is the fact that we focus on a current-current interaction [see Eq. (1)], instead of an axial current interaction. This leads to extra velocity suppressions in the nuclear bremsstrahlung rates, which combined with the fact that the neutron does not interact with $\chi$, makes bremsstrahlung subdominant for all radii except in the innermost region of the core where the positron density is very suppressed (see Fig. 9).

out and cease to affect the dark fermion flux. They are as follows:

(1) *Annihilation sphere* ($r_N$): This is the radius at which $\chi\bar{\chi} \to e^+e^-$ freezes out, or, in other words, the DM density has dropped sufficiently that the dark fermions are no longer annihilating with their antiparticles. There are effectively no number-changing reactions outside this sphere; hence it is this radius that sets the number flux of escaping DM.

(2) *Electron sphere/energy sphere* ($r_E$): This is the radius at which $\chi e \to \chi e$ freezes out. Beyond this radius, scattering events of dark fermions with electrons and positrons are no longer sufficient to keep the DM in thermal contact with the SM bath. When $r_E > r_N$, this sets the temperature of the escaping DM flux.

(3) *Streaming sphere* ($r_S$): This is the radius at which $\chi p \to \chi p$ freezes out. The proton density drops to a point that the DM is no longer diffusively trapped and the DM free-streams out of the star. Note that because the protons are significantly heavier than the DM, they cannot efficiently transfer energy to the DM; hence scattering interactions with protons do not change the energy of the DM appreciably.

In the parameter space of interest, the streaming sphere always lies well outside of the annihilation and electron spheres; hence the number flux and energy distribution are set while the DM is still diffusing. We also find that the electron sphere is always outside of the number sphere ($r_N < r_E$). As we have already discussed, even though the DM continues to scatter off of protons once outside of the electron sphere, the large discrepancy in mass between protons and the dark fermions means that the energy of the dark fermions is not largely affected during these scatterings. As a result, the energy spectrum of the DM flux is set by the temperature at $r_E$. Due to this, we will use the terms *electron sphere* and *energy sphere* interchangeably. The characteristic spheres are depicted in Fig. 1.

We can analytically compute these characteristic spheres by finding the radius at which the optical depth associated with a particular interaction becomes $\mathcal{O}(1)$ [32]. The optical depth for a given process at some radius $r_0$ is given by $\int_{r_0}^{\infty} \lambda^{-1}(r) dr$ with $\lambda(r)$ the interaction length of the process as a function of the radius. The interaction lengths for $\chi\bar{\chi}$ annihilation, $\chi e^{\pm}$ scattering, and $\chi p$ scattering are as follows:

$$\lambda_{\chi\chi}(r) = (n_\chi \sigma_{\chi\chi \to ee})^{-1}, \quad (2)$$

$$\lambda_{\chi e^{\pm}}(r) = \langle v_\chi \rangle (n_{e^{\pm}} \sigma_{\chi e \to \chi e} v_{\rm rel})^{-1}, \quad (3)$$

$$\lambda_{\chi p}(r) = (n_p \sigma_{\chi p \to \chi p})^{-1}, \quad (4)$$

with $n_X$ the number density for a species $X$, $\sigma_Y$ the cross section for a process $Y$, and $T$ the temperature of the SN at the given radius. The explicit forms of the cross sections are provided in Appendix B. Note that there is an additional factor of the averaged velocity $\langle v_\chi \rangle$ in the mean-free path for scattering off electrons. It can be understood by recalling that the rate of interactions is given by $1/\langle \sigma v_{\rm rel}\rangle$ and thus the mean-free path is $\langle v_\chi \rangle / \langle \sigma v_{\rm rel}\rangle$. For the other two interactions (annihilation and proton scattering), the relative velocity is approximately the dark matter velocity and thus these factors cancel out, but for scattering off the relativistic electrons with $v_{\rm rel} \approx 1$, this cancellation does not occur.[4]

The mean-free path for scattering with protons is much shorter than that of the other interactions in most regions of interest. This is due to the extra $v_\chi$ suppression in the scattering cross section with electrons and due to the Boltzmann suppression in $n_\chi$ for annihilation. Because of this, a DM particle will undergo many scatterings with protons in between scatterings with electrons or annihilations. This must be taken into account by using an effective optical depth for the scattering with electrons and annihilations as discussed in Ref. [32]. The characteristic radii $r_N$, $r_E$, and $r_S$ are computed using the following criteria (see, e.g., Ref. [34]):

$$\int_{r_N}^{\infty} \sqrt{\lambda_{\chi\chi}^{-1}(r)[\lambda_{\chi p}^{-1}(r) + \lambda_{\chi\chi}^{-1}(r)]} dr = \frac{2}{3}, \quad (5)$$

$$\int_{r_E}^{\infty} \sqrt{(\lambda_{\chi e^-}^{-1}(r) + \lambda_{\chi e^+}^{-1}(r))[\lambda_{\chi p}^{-1}(r) + \lambda_{\chi e^-}^{-1}(r) + \lambda_{\chi e^+}^{-1}(r)]} dr = \frac{2}{3}, \quad (6)$$

$$\int_{r_S}^{\infty} \lambda_{\chi p}^{-1}(r) dr = \frac{2}{3}. \quad (7)$$

The integrands are simply the inverse effective mean-free paths to the next relevant interaction (number-changing, energy-changing, and scattering, respectively), so the overall integral is proportional to the expected number of relevant interactions a particle will experience while escaping the protoneutron star.

Having computed these radii, one can use a similar argument to that of the neutrino sphere to make a simple estimate of the outgoing DM spectrum. The logic behind the following methodology is simply that, by definition, $r_N$ sets the total number of dark fermions that are produced (since number-changing reactions are insignificant beyond it) and $r_E$ sets the energy spectrum (since the DM is thermally decoupled beyond it). Under the assumption that the dark matter flux has reached a temporary steady state so the DM density profile is not changing in time, the flux that escapes at infinity will be set solely by the total number flux produced and the temperature at the energy sphere.

---

[4]There should also be a factor to account for the inefficiency in the energy transfer between the light electrons and the much heavier dark matter. However, there is no closed form for this factor in the mildly relativistic regime in which we are interested, and not including this factor makes the decoupling happen at larger radii (smaller temperature) which means that not including this factor results in a conservative estimate for the detection sensitivity.

The analytic estimate proceeds as follows:

(1) Treat the protoneutron star as a blackbody of radius $r_N$ with a diffusive envelope. The number flux at the blackbody surface is given by

$$\Phi_{r_N} = g_\chi \int \frac{d^3k}{(2\pi)^2} \frac{1}{e^{E_k/T_N}+1} \frac{k\cos\theta}{E_k} \Theta(\cos\theta)$$
$$= \frac{1}{2\pi^2} \int dE \frac{E^2 - m_\chi^2}{e^{E/T_N}+1}, \quad (8)$$

where $g_\chi = 4$ is the number of degrees of freedom (d.o.f.) in DM and $T_N \equiv T(r_N)$ is the temperature at the number sphere. To obtain an energy flux one can just multiply the integrand by the DM energy.

(2) Multiply this total flux by a normalized differential energy spectrum set by assuming a Fermi-Dirac distribution at $T_E$, the temperature at the energy sphere:

$$\frac{\partial \Phi_{r_E}}{\partial E} = \Phi_{r_N} \left( \frac{E^2 - m^2}{\exp(E/T_E)+1} \right)$$
$$\times \left( \int_{m_\chi}^\infty \frac{E^2 - m^2}{\exp(E/T_E)+1} dE \right)^{-1}. \quad (9)$$

(3) Even though the number changing reactions are frozen out at $r > r_N$, some particles emitted from that radius can bounce back and return to the region $r < r_N$ as they are trying to diffuse out of the streaming sphere. Therefore one must include a transmission factor to account for the losses due to this effect (see Ref. [32] for details),

$$\frac{\partial \Phi_\chi}{\partial E} = \frac{\partial \Phi_{r_E}}{\partial E} \left(1 + \frac{3}{4}\tau_{r_N}\right)^{-1}, \quad (10)$$

where $\tau_{r_N}$ is given by

$$\tau_{r_N} = \int_{r_N}^\infty dr \left(\frac{r_N}{r}\right)^2 n_p(r)$$
$$\times \int d\cos\theta (1 - \cos\theta) \frac{d\sigma_{\chi p}}{d\cos\theta}. \quad (11)$$

This approach involves a number of simplifying approximations, such as the notion of a sharp radial freeze-out for different processes, but despite its limitations it provides a simple physical picture of how the various interactions affect the DM flux. Furthermore, it serves as a cross-check on the results of the full Monte Carlo simulation. We used it as such and found that the analytic estimates agreed with simulation results to within an order of magnitude. To provide a point of comparison, we include at the end of Sec. IV C a comparison of the DM profile generated by our analytic estimate to the output of the Monte Carlo transport simulation.

## IV. FLUX COMPUTATIONS FROM SIMULATION

While the methodology outlined in the previous section provides a convenient way to understand the physics of the trapped regime, it becomes less accurate as the region over which the decoupling of a certain process occurs becomes larger (i.e., the decoupling radii become smeared into decoupling regions). To produce more robust estimates of the DM flux, one must perform a full Boltzmann particle transport simulation. To calculate the transport of dark fermions we use a Monte Carlo (MC) method that can be broken into four main steps:

(1) *Initial conditions*: To describe the underlying matter distribution of a protoneutron star we refer to detailed multiphysics dynamical simulations of core-collapse supernovae. From these we define fiducial analytic profiles that capture the temperature, density, and electron fraction structure of the cooling protoneutron star.

(2) *MC flux computation*: An iterative MC simulation is used to determine the steady-state DM distribution within the star. The resulting DM profile is used to determine the outgoing number flux of dark fermions.

(3) *Energy spectrum*: The energy distribution is set in the same manner as the analytic treatment described in the previous section. The energy sphere is computed and used to set a temperature for the outgoing flux.

(4) *Gravitational redshift*: This energy spectrum is subsequently adjusted to take into account the effects of gravitational redshift on the escaping DM.

In the following subsections, we will address each of these steps in turn.

### A. Initial conditions

The collapse of a massive star in a core-collapse supernova leads to the compression of the inner iron core into a dense protoneutron star (PNS) with a mass $M \approx 1.4\,M_\odot$ and a total thermal energy (derived from the gravitational collapse) of order $E \approx 10^{53}$ erg. At its formation, the PNS has a radius of several tens of kilometers, although over a timescale of tens of seconds it will cool and condense into a cool neutron star of radius $\sim 14$ km.

The mechanisms of core-collapse SNe and PNS formation are complex multiphysics phenomena that involve a dense matter equation of state and multidimensional effects such as turbulence and convection. While constructing high-fidelity simulations of these events remains a work in progress, the essential structure of the remnant PNS can be specified at a level appropriate for making the DM flux estimates of this paper.

The gray lines in Fig. 2 show an example PNS structure about 1 s after bounce. In the bulk interior of the PNS, the energy density is dominated by baryons, leading to temperatures of order $T \approx 30$ MeV. Deleptonization via weak interaction results in a low electron fraction $Y_e \approx 0.1$. The very central core ($r \lesssim 5$ km) of the PNS typically has a slightly higher $Y_e \approx 0.2$ and is a factor of $\sim 2$ colder, as this region is adiabatically compressed in the collapse without experiencing strong heating from shocks.

Above the PNS is a steep, hydrostatic atmosphere where the density falls off exponentially with a scale height $\lesssim 1$ km. The temperature structure in the atmosphere is roughly set by neutrino diffusion, which implies a scaling $T(r) \propto \rho(r)^{1/4}$. The neutrino sphere generally sits somewhere within this steep PNS atmosphere. The electron fraction rises, approaching symmetry ($Y_e \approx 0.5$) outside the PNS.

Finally, above the PNS atmosphere the densities drop such that radiation pressure dominates and the profile changes. The layers above the PNS are initially convective and hence nearly isentropic, which results in the density and pressure having approximately power-law profiles ($\rho(r) \propto r^{-3}, T(r) \propto r^{-1}$). Matter driven by neutrino winds from the PNS surface may also influence the structure above the atmosphere, but this still results in a similar power-law profile.

To capture these essential features of PNS structure without restricting ourselves to any specific core collapse simulation, we constructed an analytic profile that resembles the results of full simulations. Details of the analytic mapping are given in Appendix A, and a comparison of simulation data to the chosen fiducial profile is shown in Fig. 2.

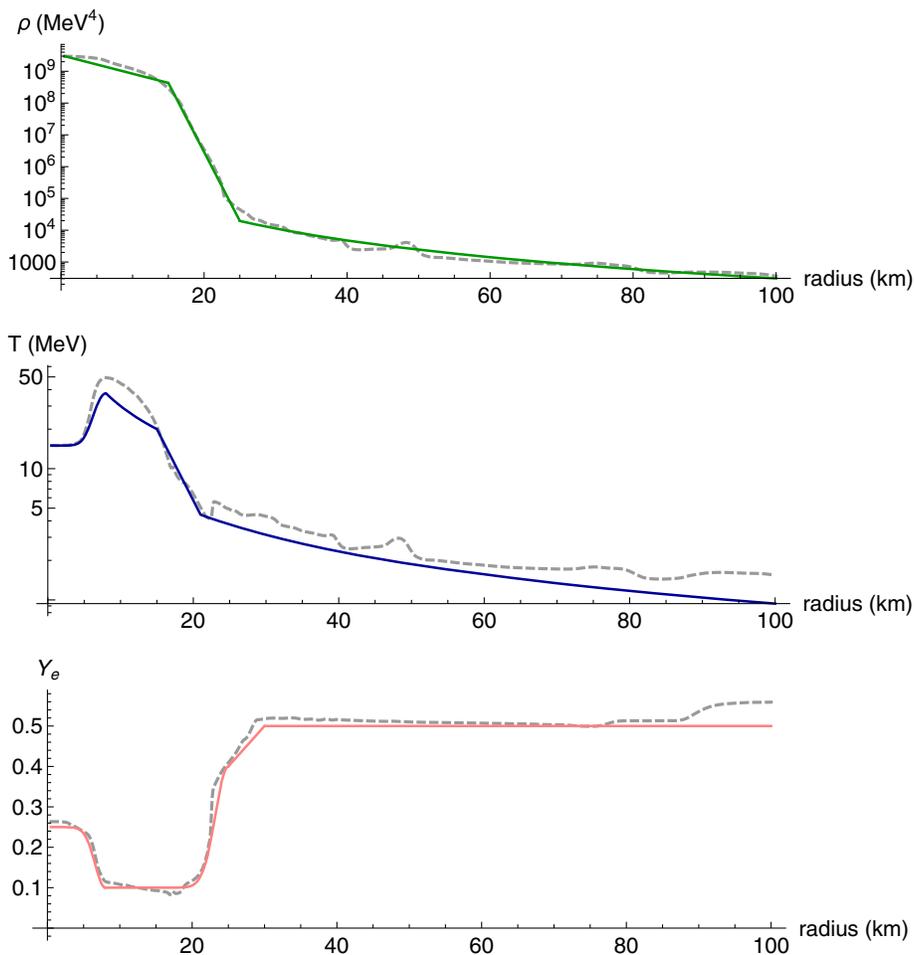

FIG. 2. The analytic profile used in this analysis (colored lines) is displayed alongside the results of one run of the supernova core-collapse simulation (dashed lines). Note the strong agreement between the analytic profiles and the simulation results. The temperature in the analytic profile is uniformly lower than simulation because it has been adjusted such that it reaches a maximum temperature of 30 MeV, which is a conservative and theoretically motivated peak core temperature (see discussion in text).

We estimated the profile dependence of our results by varying the parameters of our analytic profile. The resulting flux is most sensitive to the overall scale of the temperature since the production terms depend strongly on temperature (see Appendix B 5). Rescaling the profile such that the peak temperature changes from ∼30 MeV to ∼50 MeV results in an increase in the flux by a small $\mathcal{O}(1)$ factor for masses below ∼40 MeV and an order of magnitude for large masses. This is unsurprising given that the larger masses are already being produced on a Boltzmann tail, so the production is exponentially sensitive to the temperature. However, even order-of-magnitude changes in flux make no appreciable change to the sensitivity bounds displayed in this paper due to the fact that the flux changes very rapidly with $y$. It is true that with a higher temperature, the bounds may extend out to slightly larger masses, but we have chosen a profile with peak temperature 30 MeV so as to make our bounds conservative.

Using our analytic profiles for temperature, density, and electron fraction, it is straightforward to compute the resulting abundances of all SM particle species. To compute the proton number density, we assume that the electron and proton fractions are comparable in the protoneutron star [i.e., $Y(r) \sim Y_p(r)$] and that the protons are the dominant contribution to the total mass density. These assumptions immediately yield $n_p(r) = \frac{Y(r)\rho(r)}{m_p}$ as the proton number density.

To compute the thermal densities of the electrons and positrons, we make the assumption of thermal equilibrium and use the associated thermal abundances (see [35]),

$$n_{e^\pm}(r) = \frac{2}{2\pi^2}\int_{m_e}^{\infty}\frac{1}{\exp(\frac{E\pm\mu_e}{T})+1}E\sqrt{E^2-m_e^2}dE. \quad (12)$$

The chemical potential $\mu_e$ can be determined by enforcing charge neutrality, which requires the number density of electrons to be equal to the sum of the proton and positron number densities. This yields the following condition:

$$\frac{2}{2\pi^2}\int_{m_e}^{\infty}\left(-\frac{1}{\exp(\frac{E+\mu_e}{T})+1}+\frac{1}{\exp(\frac{E-\mu_e}{T})+1}\right)$$
$$\times E\sqrt{E^2-m_e^2}dE = n_p(r). \quad (13)$$

Critically, these profiles are assumed to be unchanged on the timescale of emission (∼10 s); hence they are maintained as a fixed background in the following step of the analysis: the MC simulation of the dark fermions.

### B. MC flux computation

Having now found the radial profiles of the SM species, we must determine the DM profile. This necessitates the use of a Monte Carlo simulation of dark fermion diffusion within the protoneutron star.

We begin by computing source and annihilation terms to be used as inputs to the simulation that dictate the DM emissivity and annihilation length, the details of which appear in Appendix B. The two primary interactions that source dark fermions are electron-positron annihilation to DM ($e^+e^- \leftrightarrow \bar{\chi}\chi$) and proton-neutron bremsstrahlung ($np \leftrightarrow np\bar{\chi}\chi$). We include both contributions, but we find that for all DM masses considered, the production from electron-positron annihilation dominates over bremsstrahlung for $r \gtrsim 5$ km corresponding to the rapid fall in electron degeneracy (see Appendix A).

Within the protoneutron star we represent the dark matter fermion field by a set of $N$ discrete tracer "packets," each of which represents a number of fermions. The initial location and energy of these DM packets are sampled randomly so as to match the total thermal DM emissivity at each location in the protoneutron star. The DM packets are propagated a distance $d$ before experiencing a matter interaction event, where $d$ is determined in standard MC fashion by $d = -\lambda \ln(z)$ where $\lambda$ is the total mean-free path and $z$ is a uniform random number between (0, 1]. If the interaction event is a scattering, the direction of the DM packet is resampled from an isotropic distribution. If the interaction event is an annihilation, the DM packet is removed from the calculation. DM particles that leave the edge of the domain are tallied as escaped.

The inclusion of self-annihilation induces a nonlinearity in the transport problem due to the fact that the DM annihilation depends on the background density field of other DM particles. We address this with an iterative approach. Initially, the DM density in each zone is assumed to be thermal at the local temperature. We then run the MC transport procedure and construct an improved DM density profile by counting the DM packets passing through each zone. The entire transport step is repeated using the newly constructed DM density profile, and this process is iterated until the density structure and emergent DM flux converges. Typically we find that ≈20 iterations is sufficient to converge to a self-consistent DM distribution. For simplicity, the annihilation cross sections are assumed to be angle and energy independent, although such effects would be straightforward to include.

We include in Fig. 3 a comparison between the analytic estimate of the DM profile described in the previous section and the results of the simulation. The simulation results are displayed in green, a purely thermal DM profile is shown in red, and a blue line shows the free-streaming behavior of our analytic profile beyond $r_S$. The profile described in the previous section is fixed to be thermal up until $r_S$ and then falls with $r^{-2}$ beyond it. It is clear that while there is an $\mathcal{O}(1)$ difference between this analytic estimate of decoupling and the simulation mainly due to the treatment of the decoupling as occurring at a single radius instead of over an entire region, the general features and scaling behavior at large radii are the same. Part of the discrepancy between the analytical estimate and the Monte Carlo result can be associated with the transmission factor in Eq. (11), which was not taken into account in the figure since it only applies to the asymptotic flux.

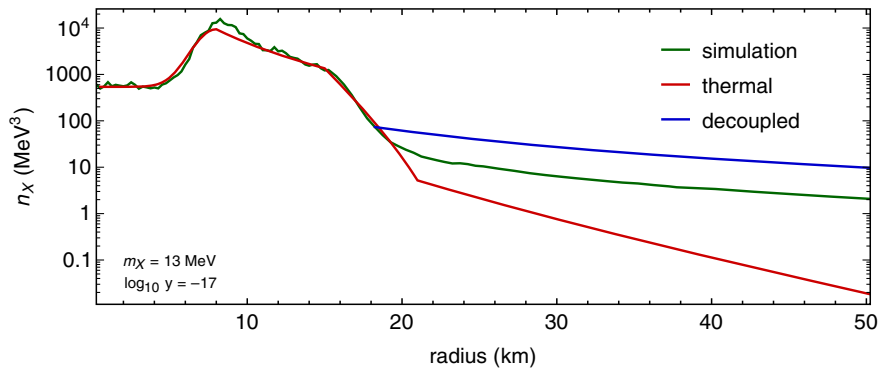

FIG. 3. Simulation results for the DM density profile for $y = 10^{-17}$ and $m_\chi = 13$ MeV are shown with a green line, with $y$ the dimensionless DM-SM coupling defined in Eq. (22). For comparison, we also display a purely thermal profile with a red line. Our analytic estimate of the profile is fixed to be thermal up to some decoupling radius, at which point, it free-streams with $r^{-2}$. This free-streaming is shown as a blue line. There is an $\mathcal{O}(1)$ discrepancy between this analytic decoupled profile and the simulation results due to the fact that the analytic profile assumes instantaneous decoupling, but the scaling behavior at large radii of the two profiles is the same.

### C. Energy spectrum

With the number flux computed from the MC simulation, we must set the energy spectrum for the escaping DM. While it would in principle be possible to extract a complete spectrum from the MC simulation itself, we find that only the dark fermions living in the high-momentum tail of the spectrum will be observable in liquid xenon detectors. Computing this tail with any precision is computationally prohibitive in that it would require the simulation to track a vast number of dark fermions such that the tail would not be dominated by statistical noise. Therefore, we instead choose to employ the analytic method detailed here to compute the spectrum because it allows for a robust prediction of the quantity of the escaping flux living in the high-momentum tail of the spectrum.

We compute the spectrum in the same manner as in the analytic methodology outlined in the previous section. Namely, we compute $r_N$ and $r_E$ using Eqs. (5)–(7), with number densities for protons, electrons, and positrons set by the abundances computed in Sec. IV A. Note that the cross sections that appear in the interaction lengths are momentum dependent. For these computations, the momentum is taken to be the average center-of-mass momentum at a given radius. This is simply $p_{\rm CM} = 3T(r)$ for DM scattering off of electrons/positrons and $p_{\rm CM} = \sqrt{6m_\chi T(r)}$ for DM scattering off of protons.

As before, we take the temperature at thermal decoupling to be $T(r_E)$. We then enforce that the DM energy spectrum take the form of a Fermi-Dirac distribution at this temperature, but with normalization set by the number flux determined via the MC simulation. Hence, we have the following differential flux:

$$\frac{\partial \dot{N}_\chi}{\partial E} = \dot{N}_\chi^{\rm MC} \left(\frac{E^2 - m^2}{\exp(E/T)+1}\right) \left(\int_{m_\chi}^\infty \frac{E^2 - m^2}{\exp(E/T)+1} dE\right)^{-1}, \quad (14)$$

where $\dot{N}_\chi = \frac{\partial N_\chi}{\partial t}$ denotes the total DM flux in number per second and $\dot{N}_\chi^{\rm MC}$ denotes the total number of DM particles escaping the PNS per second as computed with the simulation.

### D. Gravitational redshift

Finally, we must take into account the effect of gravitational redshift on the spectrum computed in the previous step. The redshifted momentum of a DM particle emitted with $p_0$ at $r_E$ is given by

$$p_\infty = p_0 \sqrt{1 - 2\Delta\Phi \left(\frac{E_0}{p_0}\right)^2} \quad (15)$$

with $\Delta\Phi$ the change in potential between $r_E$ and $r = \infty$, defined as

$$\Delta\Phi = G \int_{r_E}^\infty \frac{m_{\rm enc}(r)}{r^2} dr, \quad (16)$$

where $m_{\rm enc}(r)$ is the mass enclosed within $r$.

In the region of parameter space we are interested in, this effect does not decrease the momenta of escaping dark fermions by more than an $\mathcal{O}(1)$ factor. However, the effect does introduce a sharp cutoff in the spectrum corresponding to where the DM no longer has sufficient initial momentum to escape the gravitational well. This cutoff momentum is given by

$$p_{\min} = \sqrt{\frac{2\Delta\Phi}{1-2\Delta\Phi}m_\chi^2}. \quad (17)$$

We find that including these effects decreases the DM flux above the detector threshold by $\sim 30\%$–$40\%$.

## V. COOLING

As mentioned in the Introduction, supernovae have been used for decades to constrain models of new physics by way of a cooling argument. Our observations of the neutrino emission from SN1987a suggest a cooling timescale for the protoneutron star of $\sim 10$ s. For new d.o.f. to be compatible with this cooling timescale, they must transport energy out of the star at a rate less than the neutrinos. This simply means that new d.o.f. must transport energy out of the neutrino sphere at a rate less than $3 \times 10^{52}$ erg/s [9].

The cooling bound is usually computed more carefully in the free-streaming regime, where analytic computations can produce robust estimates of the escaping flux. However, for the trapped regime it usually relies on many approximations, and many important aspects have not been taken into account in a previous analysis. In this paper, we both extend this bound to the trapped regime using the results of our MC simulation and recompute the bound in the free-streaming regime with gravitational redshift folded in, an effect that was not included by previous papers. The upper bound and lower bound are placed in two different manners due to the fact that the upper bound (stronger couplings to the SM) will be in the trapped regime, while the lower bound (weaker couplings to the SM) will be in the free-streaming regime.

The upper bound is computed straightforwardly using the results of the simulation. The DM profiles produced by the simulation are taken to be steady-state solutions; hence the total flux going through any given radius must be constant throughout the profile. Though the cooling constraint refers to energy transport through the neutrino sphere ($\sim 20$ km), the flux of dark fermions through this radius will be equal to the flux at infinity. In all regions of parameter space that can be constrained by cooling, the energy sphere for the DM lies well within the neutrino sphere ($r_E < r_\nu$); hence we can compute the energy transfer simply by computing the fraction of the nonredshifted spectrum above $p_{\min}$ and multiplying by the flux at infinity. The cooling constraint can therefore be expressed as

$$\int_{p_{\min}}^\infty \frac{\partial \dot{N}_\chi}{\partial E}\bigg|_{E=\sqrt{p^2+m_\chi^2}} p\, dp < 3 \times 10^{52}\ \mathrm{erg/s} \quad (18)$$

with $\frac{\partial \dot{N}_\chi}{\partial E}$ defined by Eq. (14) and $p_{\min}$ defined by Eq. (17).

For the lower bound we can assume that all DM particles produced in the core will free-stream and if their velocity is above the escape velocity, they will carry energy out of the neutrino sphere. The luminosity can be calculated by the volume integral

$$L_\chi = \int_0^{r_\nu} dr\, 4\pi r^2 \left(\frac{dL_{\mathrm{brem}}}{dV} + \frac{dL_{e^+e^-}}{dV}\right), \quad (19)$$

where $dL_{\mathrm{brem}}/dV$ and $dL_{e^+e^-}/dV$ are, respectively, the local luminosities due to $np \to np\bar{\chi}\chi$ and $e^+e^- \to \bar{\chi}\chi$ in an infinitesimal volume $dV$ around a point $\vec{r}$, and $R_\nu$ is the radius of the neutrino sphere. This functions are described in Appendix D and only include particles produced with velocities above the escape velocity at a point $\vec{r}$.

## VI. DETECTION

As described in the Summary, LXe WIMP detectors are well suited to observing the high-energy dark fermion flux emitted by supernovae. It may seem at first surprising that a detector designed to detect weak-scale WIMPs would be sensitive to MeV-scale particles. Recall, however, that LXe detectors hunt for WIMPs as a constituent of the ambient galactic dark matter density. As such, the WIMPs are generally fairly cold, traveling with the galactic virial velocity of $10^{-3}$. In contrast, the dark fermions produced by SN are boosted to semirelativistic velocities, and hence have $v \sim 1$. The maximum recoil energy that an impinging DM particle with momentum $p$ could possibly deliver to a xenon nucleus is given by $\sim \frac{2p^2}{m_{\mathrm{Xe}}}$. With a WIMP of $\mathcal{O}(10)$ GeV (approximately the lower design limit for most LXe experiments) and $v = 10^{-3}$, this is a recoil energy of $\mathcal{O}(1)$ keV. Similarly, with a dark fermion of mass $\mathcal{O}(10)$ MeV and $v \sim 1$, we find a maximum recoil energy of $\mathcal{O}(1)$ keV. Unsurprisingly, given the values we chose, LXe detectors typically have thresholds on this order [27]. Since LXe detectors are already searching for WIMPs at the zero-background limit, they make for ideal targets for hunting for sub-GeV DM produced in SN.

### A. Diffuse galactic flux

It is an interesting physical consequence of the semirelativistic velocities with which the dark fermions are emitted that they will form a diffuse[5] galactic flux of energetic DM. This flux is similar to the diffuse supernova neutrino background (DSNB) (see, e.g., [36] for a review), but with the significant difference that it is due to the overlapping emissions of galactic supernovae, while the DSNB is due to extragalactic supernovae. The reason for this is that, in contrast to the neutrinos, the dark fermions are emitted traveling with an $\mathcal{O}(1)$ spread in velocity. This distribution of velocities at emission means that the DM arrives at Earth over a long period of time (comparable to the light travel time to the SN). For galactic SN, this timescale is of order $10^4$ years. With an estimated galactic SN rate of roughly 1 per century [37], we see immediately that the dark fermion emissions from up to $10^2$ galactic SN can overlap simultaneously at Earth, resulting in a diffuse galactic flux of SN-produced dark fermions.[6] (Note that since SN neutrinos are produced at $c$, they arrive in a ten-second window. It is clear that the galactic SN rate is insufficient for neutrino emissions from different SN to ever overlap; however, the extragalactic rate is suitably large enough for overlap, leading to the existence of the DSNB.)

---

[5]It should be noted that while we refer to this flux as "diffuse" because it is approximately constant in time, it is not isotropic. The flux is strongly peaked toward the galactic center (where the SN rate is highest in the galaxy). This may allow future directional detectors to discriminate this flux from a cosmological abundance of cold WIMPs. (See upcoming paper for details.)

To compute this diffuse DM flux from galactic SN, we take the double-exponential profile of Adams *et al.* [37] for the core-collapse SN density rate in our galaxy:

$$\frac{dn_{\text{SN}}}{dt} = Ae^{-r/R_d}e^{-|z|/H} \quad (20)$$

with $R$ the galactocentric radius and $z$ the height above the galactic midplane. For Type II SN, we use the parameter values Adams *et al.* provide: $R_d = 2.9$ kpc, $H = 95$ pc. Taking the galactic supernova rate to be 1 SN per 50 years, we compute $A = 0.00208$ kpc$^{-3}$ yr$^{-1}$. Earth sits at $R_E = 8.7$ kpc and $z_E = 24$ pc.

Since the flux from a given SN falls off with $1/r^2$ with $r$ the distance from the SN, we can integrate over this distribution, weighting by the $1/(\vec{r} - \vec{R}_E)^2$. The integral therefore takes the form

$$\text{total flux} = N_\chi \int_0^{z_{\max}} \int_0^{2\pi} \int_0^{R_{\max}} \frac{dn_{\text{SN}}}{dt} \frac{1}{(\vec{r} - \vec{R}_E)^2} dr d\theta dz, \quad (21)$$

where $N_\chi \equiv \dot{N}_\chi \Delta t$ is the total number of DM particles produced in a single SN over the $\Delta t = \log(10)$ second emission timescale.[7] Computing this at the Earth's location in the galaxy gives a flux on Earth of $\Phi_{\text{diffuse}} = (2.69 \times 10^{-54} \text{ cm}^{-2} \text{ s}^{-1})N_\chi$.

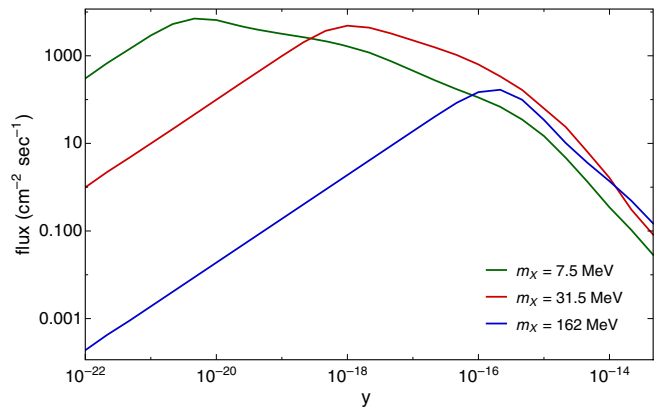

FIG. 4. We display the diffuse galactic flux of dark sector fermions on Earth as a function of $y$, the coupling to the SM, for a variety of masses. The linear portion at low couplings corresponds to the free-streaming regime, in which production scales linearly with $y$ and there is bulk volume emission of DM. At higher couplings, the PNS behaves as a blackbody and emits DM from a surface. In this trapped regime, as the coupling increases, the surface moves outwards into cooler regions and the DM flux drops accordingly.

In Fig. 4, we use the $N_\chi$ produced by our MC simulation to display the magnitude of this diffuse galactic flux on Earth as a function of $y$, a convenient variable that encapsulates the strength of the DM-SM coupling. It is defined as

$$y = \epsilon^2 \alpha_D \left(\frac{m_\chi}{m_{A'}}\right)^4 \quad (22)$$

with $\epsilon$ the small parameter controlling the kinetic mixing of the SM photon with the dark photon, $\alpha_D$ the fine-structure constant of the dark U(1) sector, and $m_{A'}$ the mass of the dark photon [24]. The free-streaming and trapped regimes are both apparent in the figure. At low couplings, the DM free-streams from the PNS and the production scales linearly with $y$; hence the flux on Earth scales linearly with $y$ as well. For larger couplings, we enter the trapped regime, where the DM is emitted from some approximately blackbody surface. As the coupling increases, this surface moves out to larger radii where the PNS is cooler; hence the DM flux decreases.

This diffuse source can be compared to the flux from a hypothetical nearby point source. We find that in order for a single SN to produce a comparable flux of DM on Earth, it would have to sit within roughly 1 kiloparsec of Earth and would have had to have occurred recently enough that the DM flux would still be passing through us. There are no observed SN that unambiguously satisfy these criteria; hence our sensitivity limits are placed using exclusively the galactic diffuse flux. However, if future observations detect such a SN, this would potentially enhance experimental sensitivity to DM flux from supernovae. Point sources and their associated recoil spectra are further discussed in Appendix C.

---

[6]The SN-produced dark fermions will also produce a diffuse extragalactic flux but in the following analysis, we conservatively ignore extragalactic contributions as they are subdominant to the galactic flux.

[7]Note that although this is a diffuse flux, the emission timescale of the SN appears simply because our analysis estimates the instantaneous DM flux from a supernova, but we need the total number of DM particles injected by a supernova in order to compute the diffuse flux from several overlapping SN. This emission timescale is set by the cooling of the PNS and is often taken to be 10 seconds, though we choose to take it to be log(10) seconds so as to be conservative.

Note that while we refer to this flux as "diffuse" because it is approximately constant in time, it is not isotropic. The flux is strongly peaked toward the galactic center (where the SN rate is highest in the galaxy). As will be discussed in upcoming work, this may allow directional detectors to discriminate this flux from a cosmological abundance of cold WIMPs.

### B. Count rates in liquid xenon detector

The final necessary piece of this analysis is to determine the detection rate of the diffuse flux in liquid xenon detectors. This is given by the following expression:

$$\text{event rate} = N_{\text{targets}} \int_{\sqrt{\frac{1}{2}m_{\text{Xe}}E_{\text{thresh}}}}^{\sqrt{\frac{1}{2}m_{\text{Xe}}E_{\text{max}}}} \int_{E_{\text{thresh}}}^{2p_\infty^2/m_{\text{Xe}}} \frac{d\sigma}{dE_{\text{rec}}}\bigg|_{p=p_\infty} \frac{d\Phi_{\text{diffuse}}}{dp_0}\bigg|_{p=p_0} dE_{\text{rec}} dp_\infty \quad (23)$$

with $\frac{d\sigma}{dE_{\text{rec}}}$ the differential DM-Xe cross section defined in Appendix B 7, $E_{\text{rec}}$ the recoil energy of the xenon nucleus, $[E_{\text{thresh}}, E_{\text{max}}]$ the recoil energies measured by the detector, and

$$\frac{d\Phi_{\text{diffuse}}}{dp_0} = \Phi_{\text{diffuse}}\left(\frac{p_0^2}{\exp\left(\frac{\sqrt{p_0^2+m_\chi^2}}{T}\right)+1}\right)\left(\frac{p_0}{\sqrt{p_0^2+m_\chi^2}}\right)\left(\int_{m_\chi}^\infty \frac{E^2-m^2}{\exp(E/T)+1}dE\right)^{-1} \quad (24)$$

the differential diffuse galactic flux of dark fermions. Note that the outer integral is taken over $p_\infty$, the dark fermion momentum at infinity [given by Eq. (15)], since the scattering is occurring on Earth; however, the factors corresponding to the energy spectrum of the DM are in terms of $p_0$, the momentum at production, since the distribution is defined at $T_{r_E}$. It is trivial to find $p_0$ by inverting Eq. (15). The limits of integration derive from requiring that the recoil energy be above threshold and less than the maximum recoil energy probed by the detector. Note that since the DM is usually very near the lower threshold for energy deposition and typical values of $E_{\text{max}}$ are usually several tens of keV [27], $E_{\text{max}}$ plays little role in determining the event rate.

In Fig. 5, we show three recoil spectra for a liquid xenon detector. We have set $\log y = -15.3$ and plot a variety of masses. All of these points lie within the interesting region of parameter space for direct detection. It is clear from the figure that lower masses result in lower average recoil energies while the tail of recoil energies can be fairly large for heavier DM owing to its larger kinetic energy. Integrating these distributions allows us to compute the number of events expected in a variety of existing and next-generation LXe detectors.

### VII. RESULTS

Our results are summarized in Figs. 6 and 7. We have chosen to display the sensitivity limits of the following detectors:

(1) *Xenon1T*: Xenon1T has already completed a one ton-year exposure with no observation of a signal above background [27]. As such, we choose to display the sensitivity region for this exposure. The xenon1T sensitivity region is shown in red.

(2) *LUX-Zeplin*: LUX-Zeplin is a LXe WIMP experiment currently under construction. When completed, it is projected to be the most sensitive LXe detector to date. It is expected to run for a total integrated exposure of 15 ton-years [30], which is the value we have used in computing our limits. Its reach is shown in yellow.

(3) *DARWIN*: DARWIN is a future LXe experiment designed to be the ultimate LXe WIMP detector,

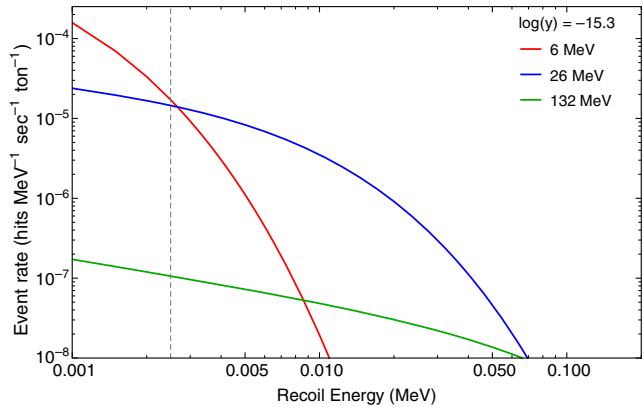

FIG. 5. Recoil spectra in a liquid xenon detector for the parameters $\log y = -15.3$ and $m_\chi$ of 6, 26, and 132 MeV. The 2.5 keV energy threshold used in many current LXe detectors is shown for reference as a dashed gray line. For fixed $y$, increasing $m_\chi$ results in lower flux, hence lower overall scaling, but a longer tail due to the larger kinetic energy of the incident DM.

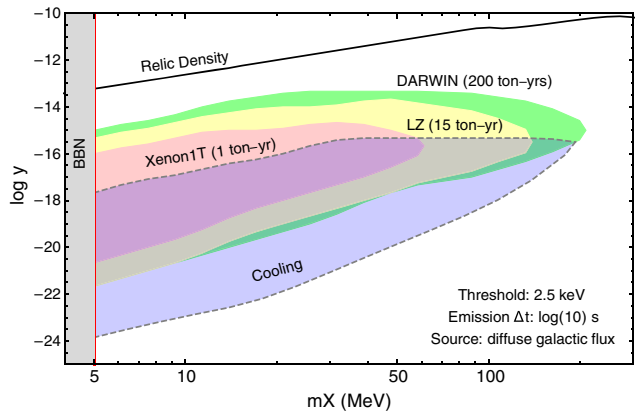

FIG. 6. The sensitivity regions for xenon1T (red line), LUX-Zeplin (yellow area), and DARWIN (green area). The detector threshold has been taken to be 2.5 keV and the emission timescale from the SN to be log 10 seconds. We compute these curves using the diffuse galactic flux. The region bounded by our cooling bound is overlaid in blue.

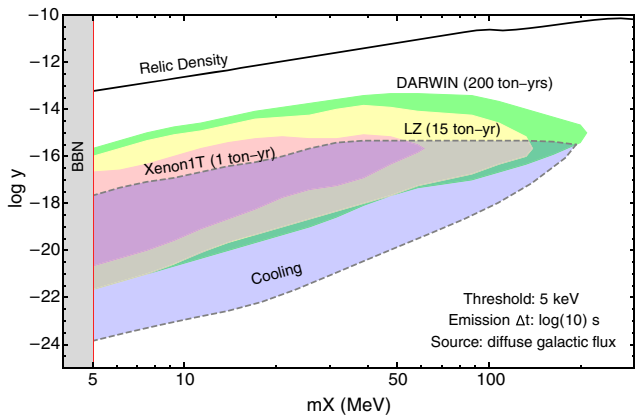

FIG. 7. Same as Fig. 6 but with detector threshold set to 5 keV. Note that this does not affect the cooling bound.

with sensitivity down to the neutrino floor [31]. If constructed, it will have an integrated exposure of 200 ton-years. Its reach is shown in red.

Existing LXe detectors generally have nuclear recoil thresholds of 5 keV [27] but future improvements aim to lower this to 2.5 keV, where solar neutrinos begin to become a large background [31]. As a result, we have chosen to display the sensitivity limits for both values. Our emission timescale has been chosen conservatively to be log(10) seconds as we do not have a precise notion of the time dependence of the profile at the radii of interest and thus assumed that the $\chi$ luminosity will decrease approximately as $1/t$ in the first 10 seconds, in analogy with the neutrino case.

The vertical axis is defined in terms of the convenient variable $y$, which is an oft-used variable in discussions of these models that serves as a measure of the coupling of the DM to the SM. Recall that $y$ is defined as $y = \epsilon^2 \alpha_D (\frac{m_\chi}{m_{A'}})^4$ with $\epsilon$ the small parameter controlling the kinetic mixing of the SM photon with the dark photon, $\alpha_D$ the fine-structure constant of the dark U(1) sector, and $m_{A'}$ the mass of the dark photon [24]. There is clearly a degeneracy between the parameters of the dark sector for a given value of $y$. It should be noted that all of the detection curves presented here are sensitivity regions, not exclusion limits. In other words, at any given point within the reach, the detector is sensitive to some choice of parameters that yields a given $y$, but is not necessarily sensitive to *all* choices of parameters. This is an important distinction given that for certain values of $\alpha_D$, the scattering of the dark fermions within the protoneutron star will be dominated by self-scattering, rather than scattering off of protons, an effect neglected in this analysis. We will treat these self-interactions in upcoming work, as well as considering models with extra structure, including a lighter dark photon and cannibalistic interactions [38,39].

The cooling region is shown in blue. The upper region is calculated in the trapped regime and is valid under our assumption that the self-interactions can be neglected. The bottom of the exclusion region is obtained from the free streaming regime and should be valid even when considering large self-interactions. Our bounds are stronger than those obtained in Ref. [20] for two main reasons: (1) their analysis only included production through nucleon-nucleon bremsstrahlung, which is subdominant in all of the parameter space we considered to the production from $e^+e^-$, and (2) their treatment of the trapped regime is more conservative in that they only consider the equivalent of the free-streaming sphere and approximate the dark matter flux as a blackbody at that radius.[8]

The relic density line is reproduced from Ref. [24] and corresponds to where the relic abundance of dark fermions produced by freeze-out matches the observed dark matter density. It is included for reference. The parameter space constrained by our analysis lies beneath this, meaning that for a standard cosmological history, the dark fermions would not have sufficient cross section to be depleted down to the measured dark matter density and thus would be overabundant. However these constraints can be avoided by considering nonstandard cosmologies with, e.g., late entropy injections or by including extra interactions in the dark sector.

## VIII. CONCLUSION

The extreme temperatures and densities that are reached during supernovae would create vast abundances of any sub-GeV d.o.f. in a dark sector. In regions of parameter space where the coupling of the dark sector to the Standard Model is too large to allow the produced dark matter to free-stream out of the cooling protonteutron star, the DM becomes diffusively trapped. In this paper, we focus on a model of an additional U(1) dark sector populated by $\mathcal{O}(1\text{--}100)$ MeV fermions and a heavy dark photon that mixes kinetically with the Standard Model photon. As the dark fermions diffuse out of the star, the flux and spectrum are set by the freeze-out of various interactions. Here, we have used this to calculate the DM flux by employing a dedicated Monte Carlo simulation of particle transport within the protonteutron star. The results allow us to extend the well-known cooling bound into the diffusive regime.

---

[8]Note that one cannot directly compare the limits displayed in their paper to those displayed here since in their analysis they specialized to the case where $m_\chi = m_A/3$ and included the production of dark photons, which leads to substantial changes compared to our analysis whenever $m_A \lesssim 200$ MeV.

In addition, the fluxes can also be sufficiently large to be detectable in existing liquid xenon WIMP detectors. Due to the semirelativistic velocities with which the fermions escape from the star, the arrival time on Earth of the flux from a single SN overlaps with $\gtrsim 10^4$ other SN, leading to a diffuse galactic flux of dark fermions permeating the Earth. We show that existing and proposed liquid xenon detectors are sensitive to this flux over a large region of parameter space. Future LXe experiments may provide the first direct detection of dark matter at the MeV scale.

Although we have focused on a particular model of such light dark matter, the same idea applies broadly to many models of DM with mass below $\sim$GeV. Existing direct detection results along with SN cooling in the trapped regime may already set important limits on these other models. Perhaps most excitingly, future direct detection experiments could very well discover a wide variety of light dark matter through supernova production.

## ACKNOWLEDGMENTS

We thank Nikita Blinov, Philip Schuster, Natalia Toro, Alex Friedland, Rouven Essig, and Sam McDermott for useful discussions. W. D., G. M. T., and P. W. G. are supported by DOE Grant No. DE-SC0012012. W. D. and P. W.G are further supported by NSF Grant No. PHY-1720397, the Heising-Simons Foundation Grants No. 2015-037 and No. 2018-0765, DOE HEP QuantISED Award No. 100495, and the Gordon and Betty Moore Foundation Grant No. GBMF7946. The work of G. M. T. was also supported by the NSF Grant No. PHY-1620074 and by the Maryland Center for Fundamental Physics. S. R. is supported in part by the NSF under Grants No. PHY-1638509 and No. PHY-1507160, the Simons Foundation Award No. 378243, and the Heising-Simons Foundation Grant No. 2015-038. D. K. is supported in part by the U.S. Department of Energy, Office of Science, Office of Nuclear Physics, under Contracts No. DE-AC02-05CH11231 and No. DE-SC0017616, by SciDAC Award No. DE-SC0018297, by the National Science Foundation Grant No. PHY-1630782, and by the Heising-Simons Foundation Grant No. 2017-228.

## APPENDIX A: ANALYTIC PROFILE OF SN

To provide an analytic fit to the results of the full multiphysics supernova simulation described in Sec. IV A, we defined a fiducial profile in the following way:

$$\rho(r) = \rho_0 \times \begin{cases} e^{-2(r-R_0)/R_0} & r < R_0 \\ e^{(R_0-r)/h} & R_0 \leq r < R_t, \\ e^{(R_0-R_t)/h}(r/R_t)^{-3} & r \geq R_t \end{cases} \quad \text{(A1)}$$

$$T(r) = \begin{cases} T_{\text{in}} + (T_0 \frac{R_0}{R_{\text{in}}} - T_{\text{in}})\exp\left[-16\frac{(r-R_{\text{in}})^2}{R_{\text{in}}^2}\right] & r < R_{\text{in}} \\ T_0\left(\frac{R_0}{r}\right) & R_{\text{in}} \leq r < R_0 \\ T_0 e^{(R_0-r)/4h} & R_0 \leq r < R_\nu \\ T_0 e^{(R_0-r)/4h}(R_\nu/r) & r \geq R_\nu \end{cases}$$
(A2)

$$Y(r) = \begin{cases} Y_{\text{in}} + (Y_0 - Y_{\text{in}})\exp\left[-16\frac{(r-R_{\text{in}})^2}{R_{\text{in}}^2}\right] & r < R_{\text{in}} \\ Y_0 + (Y_t - Y_0)\exp\left[-100\frac{(r-R_t)^2}{R_t^2}\right] & R_{\text{in}} \leq r < R_t \\ Y_t + (Y_{\text{out}} - Y_t)\frac{r-R_t}{R_{\text{out}}-R_t} & R_t \leq r < R_{\text{out}} \\ Y_{\text{out}} & r \geq R_{\text{out}} \end{cases}$$
(A3)

with the following fiducial parameters:

$$R_{\text{in}} = 8 \text{ km},$$
$$T_{\text{in}} = 15 \text{ MeV},$$
$$Y_{\text{in}} = 0.25,$$
$$R_0 = 15 \text{ km},$$
$$\rho_0 = 10^{14} \text{ g cm}^{-3},$$
$$T_0 = 20 \text{ MeV},$$
$$Y_0 = 0.1,$$
$$R_\nu = 21 \text{ km},$$
$$R_t = 25 \text{ km},$$
$$h = 1 \text{ km},$$
$$Y_t = 0.4,$$
$$R_{\text{out}} = 30 \text{ km},$$
$$Y_{\text{out}} = 0.5.$$

See Fig. 3 for a comparison of this profile to the output of the simulation. Additionally, we include in Fig. 8 the electron degeneracy normalized to the temperature as a function of radius.

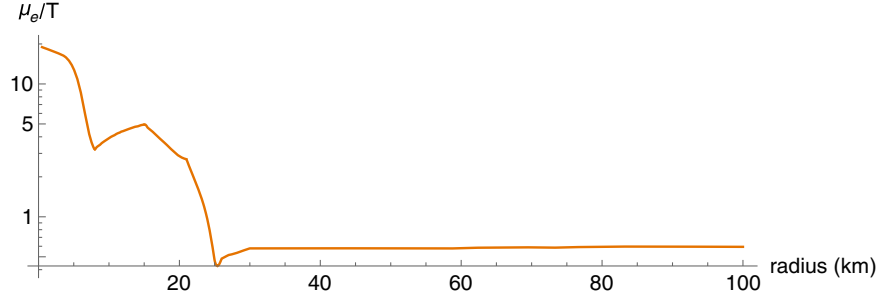

FIG. 8. The electron degeneracy divided by the temperature as a function of radius for our choice of profile.

## APPENDIX B: CROSS SECTIONS

In this appendix we list the cross sections and rates relevant for the DM dynamics in the supernova.

### 1. $\chi e \to \chi e$

This cross section is relevant for the energy decoupling of dark matter. Since for the cases of interest this is dominated at radii $\gtrsim 15$ km, we ignore the effects of Pauli blocking, which if included would decrease the cross section, leading to a smaller $r_E$ and thus to a hotter $\chi$ spectrum (and thus a more optimistic prediction for the experimental sensitivity). With this approximation the cross section in the center of momentum (COM) frame is given by

$$\sigma_{\chi e} = \frac{8\pi y \alpha}{m_\chi^2} \frac{p^2}{m_\chi^2} \left[ 1 + \frac{4}{3} \frac{p^2}{\left(p + \sqrt{p^2 + m_\chi^2}\right)^2} \right], \quad (B1)$$

where $p$ is the COM momentum and we neglected the electron mass.

### 2. $\chi p \to \chi p$

The cross section in the COM frame is given by

$$\sigma_{\chi p} = \frac{8\pi y \alpha}{m_\chi^2} \left( 2 + \frac{p^2}{m_\chi^2} \right), \quad (B2)$$

where we neglected terms that were suppressed by the proton mass.

### 3. $\bar{\chi}\chi \to e^+ e^-$

For the DM annihilation into electron-positron pairs we take Fermi blocking of the electrons into account since this is a large effect in the core, where the electron chemical potential is large. Because the cross section now depends on the electron distribution function we work in the frame of the proton-neutron star and the cross section will be in terms of the two incoming dark matter momenta $\vec{p}$ and $\vec{k}$.

First let us define the following auxiliary functions which appear frequently in the cross section due to the Pauli-blocking term

$$\begin{aligned}
B_0(E,Q,T,\mu) &= \int_{(E-Q)/2}^{(E+Q)/2} dq \left( 1 - \frac{1}{1 + e^{(q-\mu)/T}} \right) \\
&= \frac{Q}{2} + T \log\left[\cosh\left(\frac{E+Q-2\mu}{4T}\right)\right] - T \log\left[\cosh\left(\frac{E-Q-2\mu}{4T}\right)\right], \\
B_1(E,Q,T,\mu) &= \int_{(E-Q)/2}^{(E+Q)/2} dq\, q \left( 1 - \frac{1}{1 + e^{(q-\mu)/T}} \right) \\
&= \frac{T(E+Q)}{2} \log\left[1 + e^{\frac{-2\mu+E+Q}{2T}}\right] - \frac{T(E-Q)}{2} \log\left[1 + e^{\frac{-2\mu+E-Q}{2T}}\right] + T^2\left[\text{Li}_2\left(-e^{\frac{-2\mu+E+Q}{2T}}\right) - \text{Li}_2\left(-e^{\frac{-2\mu+E-Q}{2T}}\right)\right], \\
B_2(E,Q,T,\mu) &= \int_{(E-Q)/2}^{(E+Q)/2} dq\, q^2 \left( 1 - \frac{1}{1 + e^{(q-\mu)/T}} \right) \\
&= \frac{T}{4}(E+Q)^2 \log\left(1 + e^{\frac{E+Q-2\mu}{4T}}\right) - \frac{T}{4}(E-Q)^2 \log\left(1 + e^{\frac{E-Q-2\mu}{4T}}\right) \\
&\quad + T^2(E+Q)\text{Li}_2\left(-e^{\frac{-2\mu+E+Q}{2T}}\right) - T^2(E-Q)\text{Li}_2\left(-e^{\frac{-2\mu+E-Q}{2T}}\right) - 2T^3\text{Li}_3\left(-e^{\frac{-2\mu+E+Q}{2T}}\right) + 2T^3\text{Li}_3\left(-e^{\frac{-2\mu+E-Q}{2T}}\right),
\end{aligned}$$

$$(B3)$$

where $\text{Li}_n(z)$ is the Polylog of order $n$.

In order to simplify the expression we will also use the following definitions:

$$\begin{aligned}
E_p &= \sqrt{p^2 + m_\chi^2}, \\
E_k &= \sqrt{k^2 + m_\chi^2}, \\
E &= E_p + E_k, \\
\vec{Q} &= \vec{p} + \vec{k}, \\
Q &= \sqrt{p^2 + k^2 + 2pk\cos\theta},
\end{aligned} \tag{B4}$$

where $\cos\theta$ is the cosine of the angle between $\vec{p}$ and $\vec{k}$.

With those definitions, the cross section is

$$\begin{aligned}
\sigma_{\bar{\chi}\chi}(\vec{p}, \vec{k}) &= \frac{4\pi\alpha y}{m_\chi^4 \sqrt{(E_p E_k - pk\cos\theta)^2 - m_\chi^4}} \Bigg\{ 2\left[\frac{E_p^2 B_2}{Q} - \frac{E_p(\vec{p}\cdot\vec{Q})}{Q^3}[2EB_2 - (E^2 - Q^2)B_1]\right. \\
&\quad + \frac{(\vec{p}\cdot\vec{Q})^2}{2Q^5}\left[\frac{3}{4}(E^2 - Q^2)^2 B_0 - 3(E^2 - Q^2)EB_1 + (3E^2 - Q^2)B_2\right] \\
&\quad \left. + \frac{p^2}{2Q^3}\left[-\frac{(E^2 - Q^2)^2}{4}B_0 + (E^2 - Q^2)EB_1 - (E^2 - Q^2)B_2\right]\right] \\
&\quad + \frac{(E_p E_k - \vec{p}\cdot\vec{k})}{Q}[(E_p E_k - \vec{p}\cdot\vec{k}) + m_\chi^2]B_0 - 2\frac{(E_p E_k - \vec{p}\cdot\vec{k})E_p}{Q}B_1 \\
&\quad + \frac{2(E_p E_k - \vec{p}\cdot\vec{k})(\vec{p}\cdot\vec{Q})}{2Q^3}[2EB_1 - (E^2 - Q^2)B_0]\Bigg\},
\end{aligned} \tag{B5}$$

where $E$ and $\vec{Q}$ were defined in Eq. (B4) and all $B_i$ are to be interpreted as $B_i(E, Q, T, \mu)$ as defined in Eq. (B3).

### 4. Inverse bremsstrahlung annihilation term

Here we compute the DM absorption rate through inverse bremsstrahlung: $\bar{\chi}\chi np \to np$. We will use the soft radiation aApproximation (SRA), which is also used in the neutrino production (and absorption) through (inverse) bremsstrahlung [40] and also for computing dark photon production in the protoneutron star [41]. This approximation allows one to factorize the nucleon-nucleon interaction from the emission process, and the latter can be directly measured by experiment. This approximation is well justified when the energy of the emitted dark matter pair is much smaller than the COM kinetic energy of the nucleons, $\sum \omega_\chi \ll E_{\text{CM}}$. For us, this is not satisfied for most of the DM masses, and we are usually in a regime where $\sum \omega_\chi \sim E_{\text{CM}}$. In Ref. [41] it was argued that even in such a regime the SRA approximation only resulted in a factor of 2 error in the case of dark photon production. We expect that this approximation leads to an $\mathcal{O}(1)$ error in the rate, but as we will find, this rate is subdominant to the annihilation to $e^+ e^-$ almost everywhere in the proton-neutron star by a significant margin.

The absorption rate for DM via inverse bremsstrahlung is given by

$$\Gamma_\chi = \frac{1}{n_\chi} \int \frac{d^3k_1 d^3k_2}{(2\pi)^6 4\omega_1\omega_2} g(k_1)g(k_2) \int \frac{d^3p_1 \cdots d^3p_4}{(2\pi)^{12} 2E_1 \cdots 2E_4} (2\pi)^4 \delta^4(k_1 + k_2 + p_1 + p_2 - p_3 - p_4) \times f_p(p_1)f_n(p_2)|\bar{\mathcal{M}}|^2_{\bar{\chi}\chi np}, \tag{B6}$$

where $g(k)$ is the distribution function for DM (including the number of spin dof), $n_\chi$ is the number density of DM, $f_{p/n}$ is the distribution function for protons/neutrons, and $|\bar{\mathcal{M}}|^2_{\bar{\chi}\chi np}$ is the averaged matrix element squared for the $\bar{\chi}\chi pn \to pn$ process. Now, using SRA, we can rewrite this as

$$\Gamma_\chi = \frac{1}{n_\chi} \int \frac{d^3k_1 d^3k_2}{(2\pi)^6 4\omega_1\omega_2} g(k_1)g(k_2) \int \frac{d^3p_1 \cdots d^3p_4}{(2\pi)^{12} 2E_1 \cdots 2E_4} (2\pi)^4 \delta^4(p_1+p_2-p_3-p_4)$$
$$\times f_p(p_1)f_n(p_2)|\bar{\mathcal{M}}|^2_{np} \sum_{\text{spins}} \frac{1}{4} \left(\frac{eg_d\epsilon}{m_{A'}^2}\right)^2 |J^\mu \bar{u}_{k_1}\gamma_\mu v_{k_2}|^2, \tag{B7}$$

where

$$J^\mu = \frac{p_1^\mu}{p_1 \cdot k} - \frac{p_3^\mu}{p_3 \cdot k}, \qquad k^\mu = k_1^\mu + k_2^\mu, \tag{B8}$$

with $p_{1(3)}$ the momentum of the incoming (outgoing) proton and the sum over spin in the previous equation being over the DM spin.

Note that in this approximation we drop the momentum of DM in the energy momentum conservation delta function, since in the SRA these are soft compared to the nucleon energy and momentum. Because of this, we can first perform the $d^3k_i$ integrals. For this it is useful to first compute

$$R_1 = \int \frac{d\Omega_{k_1} d\Omega_{k_2}}{(4\pi)^2} J_\mu J_\nu \mathrm{tr}[(\slashed{k}_1 - m_\chi)\gamma^\mu(\slashed{k}_2 + m_\chi)\gamma^\nu]. \tag{B9}$$

We can make use of the SRA and also of the nonrelativistic (NR) nature of the nuclei to expand $J_\mu$ as a series in the nuclei velocity to lowest order. With this we find

$$R_1 = \frac{4\Delta \vec{p}^2}{M^2 \omega^4} \left[ -\frac{2}{3}(\omega_1^2 k_2^2 + \omega_2^2 k_1^2) - (\omega_1\omega_2 + m_\chi^2)\left(\frac{k_1^2+k_2^2}{3} - \omega^2\right) + \frac{2k_1^2 k_2^2}{9} \right], \tag{B10}$$

where $\omega = \omega_1 + \omega_2$ and $m_p$ is the mass of the nuclei.

We can now rewrite Eq. (B7) as

$$\Gamma_\chi = \frac{1}{n_\chi} \frac{16\pi^2 \alpha y}{m_\chi^4} R_{pn} R_\chi,$$
$$R_\chi = \int \frac{d^3k_1 d^3k_2}{(2\pi)^6 4\omega_1\omega_2} g(\omega_1)g(\omega_2) \frac{1}{m_p^2 \omega^4} \left[(\omega_1\omega_2 + m_\chi^2)\left(\omega^2 - \frac{k_1^2+k_2^2}{3}\right) + \frac{2k_1^2 k_2^2}{9} - \frac{2}{3}(\omega_1^2 k_2^2 + \omega_2^2 k_1^2)\right],$$
$$R_{pn} = \int \frac{d^3p_1 \cdots d^3p_4}{(2\pi)^{12} 2E_1 \cdots 2E_4} (2\pi)^4 \delta^4(p_1+p_2-p_3-p_4) f_p(p_1) f_n(p_2) |\bar{\mathcal{M}}|^2_{np}(\vec{p}_1-\vec{p}_3)^2. \tag{B11}$$

As a first step to compute $R_{pn}$ we first compute

$$\Pi_{pn} = \int \frac{d^3p_3 d^3p_4}{(2\pi)^6 2E_3 2E_4} (2\pi)^4 \delta^4(p_1+p_2-p_3-p_4)(p_1^2+p_3^2-2p_1 p_3 \cos\theta) 64\pi^2 E_{cm}^2 \frac{d\sigma_{np}}{d\Omega_{CM}},$$
$$\Pi_{pn} = 16 p_{CM}^3 m_p \int d\Omega_{CM}(1-\cos\theta_{CM}) \frac{d\sigma_{np}}{d\Omega}\bigg|_{CM} \equiv 16 p_{CM}^3 m_p \langle \sigma_{np}^{(2)} \rangle. \tag{B12}$$

The integral in the above expression has been obtained from the measured phase shifts in [41] and is denoted by $\langle \sigma_{np}^{(2)} \rangle$. Note that this is a function only of the CM momentum.

Using the NR approximation for the nuclei we can write $R_{pn}$ as

$$R_{pn} = \frac{n_p n_\chi}{(2\pi m_p T)^3} \int \frac{d^3 p_1 d^3 p_2}{4m_p^2} \exp\left(\frac{-p_1^2 - p_2^2}{2m_p T}\right) 16 m_p \left(\frac{|\vec{p}_1 - \vec{p}_2|}{2}\right)^3 \langle \sigma_{np}^{(2)} \rangle,$$

$$\vec{q} = \frac{\vec{p}_1 - \vec{p}_2}{2}, \qquad \vec{P} = \vec{p}_1 + \vec{p}_3,$$

$$R_{pn} = \frac{n_p n_n}{(2\pi m_p T)^3} \frac{4}{m_p} \int d^3 P \exp(-P^2/4m_p T) \int d^3 q\, q^3 \langle \sigma_{np}^{(2)} \rangle \exp(-q^2/m_p T)$$

$$= \frac{n_p n_n}{(\pi m_p T)^{3/2}} 8\pi m_p^2 \int dK\, K^2 \langle \sigma_{np}^{(2)} \rangle \exp(-K/T). \tag{B13}$$

Combining these results we find

$$\Gamma_\chi = \frac{64\alpha y}{9\pi} \frac{n_p n_n}{n_\chi} \frac{1}{(\pi m_p T)^{3/2}} \int_1^\infty dx_1 \int_1^\infty dx_2 \frac{\sqrt{(x_1^2 - 1)(x_2^2 - 1)}}{(e^{mx_1/T} + 1)(e^{mx_2/T} + 1)(x_1 + x_2)^4}$$

$$\times [4 + x_1 x_2 (3x_1^2 + 4x_1 x_2 + 3x_2^2) + (5x_1^2 + 12 x_1 x_2 + 5x_2^2)] \int_0^\infty dK\, K^2 \langle \sigma_{np}^{(2)} \rangle e^{-K/T}. \tag{B14}$$

### 5. Source term: $e^+ e^-$ channel

The source term can directly be calculated from the annihilation term by using detailed balance:

$$S_{e^+ e^-} = \int \frac{d^3 q\, d^3 q'}{(2\pi)^6 2E_q 2E_{q'}} \frac{4}{(e^{(E_q - \mu_e)/T} + 1)(e^{(E_{q'} + \mu_e)/T} + 1)}$$

$$\times \int \frac{d^3 k\, d^3 k'}{(2\pi)^6 2\omega_k 2\omega_{k'}} (2\pi)^4 \delta^4(k + k' - q - q') |\mathcal{M}|^2 (1 - g(k))(1 - g(k'))$$

$$= \int \frac{d^3 k\, d^3 k'}{(2\pi)^6 2\omega_k 2\omega_{k'}} \frac{4\sqrt{(\omega_k \omega_{k'} - kk' \cos\theta)^2 - m_\chi^4}}{(e^{\omega_k/T} + 1)(e^{\omega_{k'}/T} + 1)} \sigma_{\bar\chi\chi}(\vec{k}, \vec{k}'), \tag{B15}$$

where $\mu_e$ is the electron chemical potential, $k$ ($k'$) is the $\chi$ ($\bar\chi$) momentum, $g(k)$ is the thermal DM distribution function (per d.o.f.), and $\sigma_{\bar\chi\chi}$ is the DM annihilation cross section to $e^+ e^-$, including Fermi blocking, as defined in Eq. (B5).

### 6. Source term: Bremsstrahlung channel

We can also compute this term by enforcing detailed balance and recycle the result from Eq. (B14)[9]

$$S_{\text{brem}} = \int \frac{d^3 p_1 \cdots d^3 p_4}{(2\pi)^{12} 2E_1 \cdots 2E_4} \int \frac{d^3 k_1 d^3 k_2}{(2\pi)^6 2\omega_1 2\omega_2} (2\pi)^4 \delta^4(p_1 + p_2 - p_3 - p_4 - k_1 - k_2) \times f_p(p_1) f_n(p_2) |\mathcal{M}|^2_{\text{prod}}$$

$$= \int \frac{d^3 p_1 \cdots d^3 p_4}{(2\pi)^{12} 2E_1 \cdots 2E_4} \int \frac{d^3 k_1 d^3 k_2}{(2\pi)^6 2\omega_1 2\omega_2} (2\pi)^4 \delta^{(4)}(p_1 + p_2 + k_1 + k_2 - p_3 - p_4) \times |\bar{\mathcal{M}}_{\text{abs}}|^2 f_p(p_1) f_n(p_2) g(k_1) g(k_2)$$

$$= n_\chi \Gamma_\chi, \tag{B16}$$

where $\mathcal{M}_{\text{prod}}$ and $\mathcal{M}_{\text{abs}}$ are the production and absorption matrix elements and $g(k_i)$ are the thermal distribution functions for DM. Note that bremsstrahlung production is subdominant to $e^+ e^-$ production the full parameter space of interest. See Fig. 9.

---

[9]We have also independently computed the source term by directly using the SRA for production. We have found that both answers disagree by more than an order of magnitude for most of the masses of interest, with the direct production rate always being larger than what was obtained by enforcing a detailed balance. This effect is related to the failure of the SRA in the regime of interest, and we chose to enforce detailed balance in order to ensure that in the large coupling regime the distribution of DM would approach a thermal distribution.

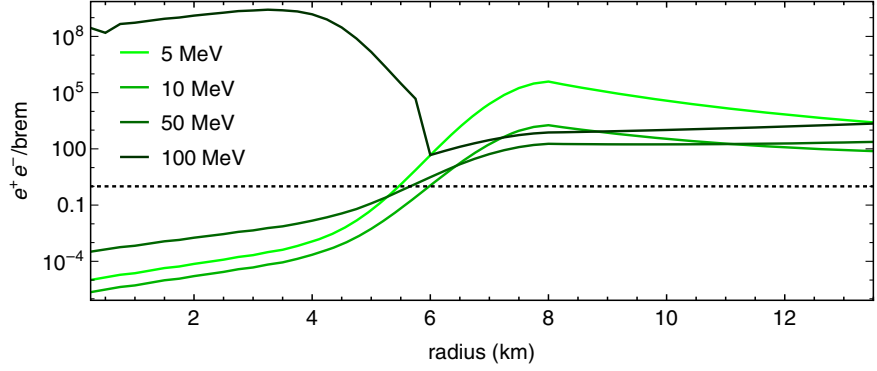

FIG. 9. This plot displays the $e^+e^-$ source term divided by the bremsstrahlung source term as a function of radius for a variety of masses. It is clear that only in the inner core does bremsstrahlung dominate for any mass. Note that the behavior of the 100 MeV curve is due to the fact that for radii less than 5 km, we have $T \ll 2m_\chi$; hence there is a large Boltzmann suppression for the bremsstrahlung. However, the chemical potential remains well above the mass ($\mu \gg m_\chi$), and hence there is little suppression on the $e^+e^-$ channel.

### 7. DM-xenon recoil

Here we compute the differential cross section for DM colliding with a xenon nucleus. Since $m_{Xe} \gg m_\chi$, we make the approximation that the center-of-mass frame and rest frame for the nucleus are roughly the same. Then, we can compute the scattering cross section and make the requisite substitutions in order to solve for it in terms of the incoming momentum of the DM particle. This gives the expression

$$\frac{d\sigma}{dE_{\rm rec}} = 4\pi\alpha y Z^2 \left(\frac{m_{Xe}}{p^2}\right) \left(\frac{1}{m_\chi^4 \left(\sqrt{m_{Xe}^2 + p^2} + \sqrt{m_\chi^2 + p^2}\right)^2}\right)$$
$$\times \left[p^2\left(1 - \frac{m_{Xe} E_{\rm rec}}{p^2}\right)\left(2\sqrt{m_{Xe}^2 + p^2}\sqrt{m_\chi^2 + p^2} + m_{Xe}^2 + m_\chi^2\right)\right.$$
$$\left. + p^2\left(2\sqrt{m_{Xe}^2 + p^2}\sqrt{m_\chi^2 + p^2} + m_\chi^2 + 3p^2\right) + m_{Xe}^2(2m_\chi^2 + p^2) + p^4\left(1 - \frac{m_{Xe} E_{\rm rec}}{p^2}\right)^2\right], \quad {\rm (B17)}$$

where $p$ is the momentum of the incoming DM particle and $E_{\rm rec}$ is the recoil energy of the xenon nucleus. The nuclear charge is $Z = 54$ for xenon.

### APPENDIX C: RECOIL SPECTRA FROM NEARBY SN

Though there are no observed supernovae that would produce a singular flux in excess of the diffuse SN background of dark fermions discussed in the body of this paper, it is still interesting to consider the recoil spectrum from a single point source. Since the fermions are produced with an $\mathcal{O}(1)$ spread in velocities, the arrival time varies between different parts of the spectrum. Dark fermions living on the high-energy (high-velocity) end of the spectrum will arrive far sooner than those on the low-energy (hence low-velocity) end. The majority of the flux will arrive with a delay behind the neutrinos of order the light-travel time to the SN.

As a result of this, the recoil spectrum of xenon in a liquid xenon detector on Earth changes over time. Shortly after the arrival of light from the SN, we expect to see a recoil spectrum that extends to high recoil energies (due to the highly boosted fermions) but with low event rates (due to the fact that the high-velocity fermions live on a tail of the spectrum). As time passes, event rates will increase but the average recoil energy will decrease as the more abundant, less energetic part of the dark fermion distribution begins to arrive on Earth. This evolution is displayed in Fig. 10. For the purposes of computation, we have focused on the case of a 30 MeV dark fermion with $\log y = -16.3$ and an SN occurring 30 kpc from Earth (the distance to the galactic center). The recoil spectra are plotted for three different time delays: $10^3$, $10^4$, and $10^5$ years after the arrival of the neutrinos on Earth. As expected, the shortest time delay corresponds to the highest energies of dark fermions; hence we have a relatively low yield, but energetic recoil spectrum. As we move toward longer delays, the average recoil energy decreases, but the event rate increases. At $10^5$ years (the light-travel time for 30 kpc), we reach the maximal event rate since this corresponds to the arrival of the peak of the dark fermion spectrum. By $10^6$ years (not shown), the dark fermion flux is once again very low since it corresponds to the arrival of the low-energy tail. The average recoil energy is well below the detector threshold.

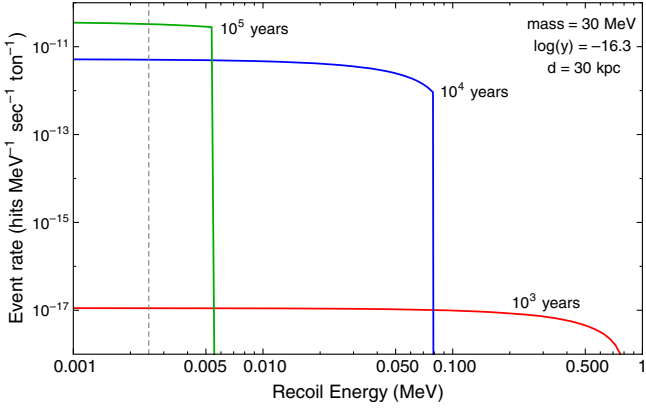

FIG. 10. The recoil spectra of xenon nuclei in a liquid xenon detector plotted for different time delays from a nearby SN. The curves shown here are for a 30 MeV dark fermion with $\log y = -16.3$ and Earth-SN distance of 30 kpc. Note the evolution of the spectrum with time, changing from an energetic spectrum with a low event rate during the arrival of the high-momentum end of the DM spectrum to a less-energetic spectrum with higher yields as the bulk of the DM spectrum arrives on Earth. The gray line indicates the 2.5 keV threshold of future LXe experiments.

We find this change in recoil spectrum a noteworthy feature of the SN flux as it could provide a discriminatory tool for detecting a DM flux from a future nearby SN, and we have included it for completeness.

## APPENDIX D: COOLING IN THE FREE-STREAMING REGIME

The lower limits of the cooling bound in Figs. 6 and 7 are obtained by considering the free-streaming regime of DM produced in the SN. In this case all DM produced in the core can free-stream out of the SN as long as it has enough kinetic energy to escape the gravitational attraction due to the protoneutron star.

In order to compute the minimum escape energy from a region of radius $r$ we need to compute the metric inside the protoneutron star. Following Ref. [42], the metric can be written as

$$ds^2 = B(r)dt^2 - A(r)dr^2 - r^2 d\Omega^2. \quad (D1)$$

The two functions $A$ and $B$ are given by

$$A(r) = \left[1 - \frac{2GM(r)}{r}\right]^{-1}, \quad (D2)$$

where

$$M(r) = \int_0^r dr' 4\pi r'^2 \rho(r') \quad (D3)$$

and

$$B(r) = \exp\left[-\int_r^\infty dr' \frac{2G}{r'^2}(M(r') + 4\pi r'^3 p(r'))\right.$$
$$\left. \times \left(1 - \frac{2GM(r')}{r'}\right)^{-1}\right], \quad (D4)$$

with $p$ the pressure in the star. Given that the pressure term is subdominant, we can approximate the pressure by treating the protons and neutrons as a gas of degenerate fermions to the level of precision we are interested in. The minimum energy required for DM to escape from a radius $r$ is given by

$$E_{\text{esc}} = \frac{m_\chi}{\sqrt{B(r)}}. \quad (D5)$$

As discussed in previous sections there are two important production channels which contribute to the DM production: electron-positron annihilation to DM and DM bremsstrahlung from proton-neutron scattering. For the profile used in our work we found that the latter yields a larger production for all masses of interest, but we include both contributions for completeness.

For the bremsstrahlung case, we use a similar calculation to what was done in Sec. B 4. However, because we are now interested in the energy flux and not the number flux, and because we must impose a minimum energy due to gravitational trapping, we cannot utilize that result which was obtained via detailed balance. The steps to compute the production cross section are almost identical, except that one must impose a maximum energy cutoff for the DM produced by hand, since due to the SRA the energy of the DM no longer appears in the energy conserving delta function. For that purpose we include an exponential regulator $\exp[-(\omega_1 + \omega_2)/T]$, where $\omega_i$ is the DM energy and $T$ the temperature.[10] Using this, the local DM luminosity from this channel is given by

$$\frac{dL_{\text{brem}}}{dV} = \frac{64\alpha y}{9\pi} \frac{n_p n_n}{(\pi m_p T)^{3/2}} \int_0^\infty K^2 \langle \sigma_{np}^{(2)} \rangle e^{-K/T} m_\chi \left[ \int_{1/\sqrt{B}}^\infty dx_1 \int_{1/\sqrt{B}}^\infty dx_2 (x_1 + x_2) \right.$$
$$+ 2 \int_{1/\sqrt{B}}^\infty dx_1 \int_1^{1/\sqrt{B}} dx_2 x_1 \left] \left( \frac{\sqrt{(x_1^2-1)(x_2^2-1)}}{(x_1+x_2)^4} \right) [4 + x_1 x_2 (3x_1^2 + 4x_1 x_2 + 3x_2^2)$$
$$+ (5x_1^2 + 12 x_1 x_2 + 5x_2^2)], \tag{D6}$$

where the first integral over $dx_i$ corresponds when both pair-produced DM have energy above the escape energy $m_\chi/\sqrt{B}$ and the second one when only one of them does.

---

[10]Another option is to introduce a hard cutoff on the DM energy such that $\omega_1 + \omega_2 \leq |\vec{p}_1 - \vec{p}_2|^2/(4m_p)$, where $\vec{p}_{1(2)}$ are the nuclei initial momentum. This guarantees that the produced energy is smaller than the COM kinetic energy of the nuclei. We found that the exponential regulator gives a smaller (and thus more conservative) rate, and also that it gives an answer that is closer to satisfying detailed balance when compared to the absorption rate.

For the electron-positron annihilation term the full form of the production above a certain energy threshold is very complicated due to the average over the initial electron and positron momentum. In order to simplify our treatment we compute the luminosity such that half the COM energy $\sqrt{s}/2$ is above $m_\chi/\sqrt{B}$ and consider that $\sqrt{s}/2$ of energy is carried away (i.e., we only consider the energy carried by the particle which gains energy from the boost from the COM frame to the star frame). Since we do not include the enhancement of the energy due to the boost, and only consider one of the produced DM particles for the luminosity, this leads to a conservative estimate. Using this the luminosity from electron positron annihilation is given by

$$\frac{dL_{e^+e^-}}{dV} = \frac{4\alpha y}{3\pi^3 m_\chi^4} \int \frac{d\omega_1 d\omega_2 \omega_1^2 \omega_2^2}{(e^{(\omega_1+\mu)/T}+1)(e^{(\omega_1-\mu)/T}+1)} \Theta(\omega_1 \omega_2 - m_\chi^2/B)$$
$$\times \int_{-1}^{1 - \frac{2m_\chi^2}{\omega_1 \omega_2 B}} d\cos\theta \sqrt{\frac{\omega_1\omega_2(1-\cos\theta)}{2} - m_\chi^2} [2\omega_1 \omega_2 (1 - \cos\theta) + 2m_\chi^2], \tag{D7}$$

where the $\Theta$ ensures that the COM energy is above the escape energy.

## APPENDIX E: TABLES

This appendix contains the full tables used to generate our bounds.

### 1. Number flux

The following tables show the number flux of DM from the SN, as determined by the MC simulation. It is the total number of DM particles escaping the SN per second.

| $\log_{10} y$ | $m_X = 5$ MeV | 6 MeV | 8 MeV | 9 MeV | 11 MeV | 14 MeV | 17 MeV | 21 MeV | 26 MeV | 32 MeV |
|---|---|---|---|---|---|---|---|---|---|---|
| −13.3 | $2.3 \times 10^{52}$ | $2.9 \times 10^{52}$ | $5.9 \times 10^{52}$ | $4.0 \times 10^{52}$ | $6.2 \times 10^{52}$ | $1.0 \times 10^{53}$ | $9.6 \times 10^{52}$ | $1.4 \times 10^{53}$ | $1.5 \times 10^{53}$ | $1.7 \times 10^{53}$ |
| −13.7 | $1.0 \times 10^{53}$ | $1.7 \times 10^{53}$ | $2.1 \times 10^{53}$ | $2.1 \times 10^{53}$ | $2.7 \times 10^{53}$ | $4.0 \times 10^{53}$ | $4.5 \times 10^{53}$ | $5.5 \times 10^{53}$ | $5.5 \times 10^{53}$ | $3.8 \times 10^{53}$ |
| −14.0 | $5.0 \times 10^{53}$ | $4.6 \times 10^{53}$ | $6.6 \times 10^{53}$ | $8.9 \times 10^{53}$ | $1.1 \times 10^{54}$ | $1.1 \times 10^{54}$ | $2.2 \times 10^{54}$ | $2.1 \times 10^{54}$ | $2.3 \times 10^{54}$ | $5.3 \times 10^{54}$ |
| −14.3 | $2.3 \times 10^{54}$ | $2.5 \times 10^{54}$ | $2.9 \times 10^{54}$ | $4.0 \times 10^{54}$ | $3.4 \times 10^{54}$ | $4.3 \times 10^{54}$ | $7.0 \times 10^{54}$ | $7.3 \times 10^{54}$ | $1.7 \times 10^{55}$ | $1.1 \times 10^{55}$ |
| −14.7 | $8.9 \times 10^{54}$ | $1.1 \times 10^{55}$ | $9.7 \times 10^{54}$ | $1.2 \times 10^{55}$ | $1.4 \times 10^{55}$ | $2.6 \times 10^{55}$ | $2.3 \times 10^{55}$ | $1.7 \times 10^{55}$ | $4.6 \times 10^{55}$ | $6.5 \times 10^{55}$ |
| −15.0 | $2.4 \times 10^{55}$ | $2.7 \times 10^{55}$ | $3.6 \times 10^{55}$ | $3.8 \times 10^{55}$ | $5.3 \times 10^{55}$ | $6.4 \times 10^{55}$ | $8.0 \times 10^{55}$ | $1.1 \times 10^{56}$ | $1.1 \times 10^{56}$ | $1.1 \times 10^{56}$ |
| −15.3 | $5.8 \times 10^{55}$ | $7.3 \times 10^{55}$ | $7.6 \times 10^{55}$ | $1.0 \times 10^{56}$ | $1.2 \times 10^{56}$ | $1.5 \times 10^{56}$ | $2.0 \times 10^{56}$ | $2.6 \times 10^{56}$ | $3.3 \times 10^{56}$ | $4.2 \times 10^{56}$ |
| −15.7 | $1.2 \times 10^{56}$ | $1.2 \times 10^{56}$ | $1.5 \times 10^{56}$ | $1.7 \times 10^{56}$ | $1.8 \times 10^{56}$ | $2.0 \times 10^{56}$ | $3.2 \times 10^{56}$ | $4.7 \times 10^{56}$ | $6.1 \times 10^{56}$ | $6.8 \times 10^{56}$ |
| −16.0 | $1.7 \times 10^{56}$ | $2.0 \times 10^{56}$ | $2.3 \times 10^{56}$ | $2.4 \times 10^{56}$ | $2.9 \times 10^{56}$ | $4.8 \times 10^{56}$ | $5.5 \times 10^{56}$ | $9.4 \times 10^{56}$ | $1.2 \times 10^{57}$ | $1.4 \times 10^{57}$ |
| −16.3 | $3.1 \times 10^{56}$ | $2.7 \times 10^{56}$ | $3.4 \times 10^{56}$ | $3.9 \times 10^{56}$ | $5.3 \times 10^{56}$ | $7.4 \times 10^{56}$ | $1.2 \times 10^{57}$ | $1.3 \times 10^{57}$ | $1.8 \times 10^{57}$ | $2.3 \times 10^{57}$ |
| −16.7 | $3.5 \times 10^{56}$ | $4.2 \times 10^{56}$ | $5.6 \times 10^{56}$ | $6.5 \times 10^{56}$ | $7.7 \times 10^{56}$ | $1.5 \times 10^{57}$ | $1.9 \times 10^{57}$ | $2.3 \times 10^{57}$ | $2.9 \times 10^{57}$ | $3.2 \times 10^{57}$ |
| −17.0 | $5.6 \times 10^{56}$ | $6.9 \times 10^{56}$ | $9.4 \times 10^{56}$ | $1.3 \times 10^{57}$ | $1.6 \times 10^{57}$ | $2.0 \times 10^{57}$ | $2.8 \times 10^{57}$ | $3.3 \times 10^{57}$ | $4.0 \times 10^{57}$ | $4.6 \times 10^{57}$ |
| −17.3 | $8.3 \times 10^{56}$ | $1.2 \times 10^{57}$ | $1.5 \times 10^{57}$ | $2.0 \times 10^{57}$ | $2.3 \times 10^{57}$ | $3.4 \times 10^{57}$ | $3.9 \times 10^{57}$ | $4.2 \times 10^{57}$ | $4.8 \times 10^{57}$ | $6.7 \times 10^{57}$ |
| −17.7 | $1.3 \times 10^{57}$ | $1.9 \times 10^{57}$ | $2.7 \times 10^{57}$ | $3.1 \times 10^{57}$ | $3.7 \times 10^{57}$ | $4.4 \times 10^{57}$ | $5.1 \times 10^{57}$ | $5.4 \times 10^{57}$ | $6.7 \times 10^{57}$ | $9.7 \times 10^{57}$ |
| −18.0 | $1.8 \times 10^{57}$ | $2.7 \times 10^{57}$ | $3.4 \times 10^{57}$ | $4.3 \times 10^{57}$ | $5.0 \times 10^{57}$ | $5.6 \times 10^{57}$ | $6.5 \times 10^{57}$ | $7.5 \times 10^{57}$ | $1.1 \times 10^{58}$ | $1.2 \times 10^{58}$ |
| −18.3 | $2.4 \times 10^{57}$ | $3.2 \times 10^{57}$ | $4.5 \times 10^{57}$ | $5.5 \times 10^{57}$ | $6.3 \times 10^{57}$ | $7.8 \times 10^{57}$ | $8.5 \times 10^{57}$ | $1.2 \times 10^{58}$ | $1.3 \times 10^{58}$ | $8.9 \times 10^{57}$ |
| −18.7 | $3.4 \times 10^{57}$ | $4.3 \times 10^{57}$ | $5.5 \times 10^{57}$ | $6.6 \times 10^{57}$ | $8.5 \times 10^{57}$ | $9.9 \times 10^{57}$ | $1.3 \times 10^{58}$ | $1.5 \times 10^{58}$ | $9.5 \times 10^{57}$ | $4.4 \times 10^{57}$ |
| −19.0 | $4.4 \times 10^{57}$ | $5.5 \times 10^{57}$ | $6.6 \times 10^{57}$ | $7.8 \times 10^{57}$ | $1.1 \times 10^{58}$ | $1.4 \times 10^{58}$ | $1.6 \times 10^{58}$ | $1.0 \times 10^{58}$ | $4.6 \times 10^{57}$ | $2.0 \times 10^{57}$ |
| −19.3 | $5.6 \times 10^{57}$ | $6.6 \times 10^{57}$ | $7.7 \times 10^{57}$ | $1.0 \times 10^{58}$ | $1.4 \times 10^{58}$ | $1.6 \times 10^{58}$ | $1.0 \times 10^{58}$ | $4.9 \times 10^{57}$ | $2.1 \times 10^{57}$ | $9.4 \times 10^{56}$ |
| −19.7 | $6.6 \times 10^{57}$ | $7.4 \times 10^{57}$ | $9.4 \times 10^{57}$ | $1.5 \times 10^{58}$ | $1.7 \times 10^{58}$ | $1.1 \times 10^{58}$ | $5.1 \times 10^{57}$ | $2.3 \times 10^{57}$ | $1.0 \times 10^{57}$ | $4.4 \times 10^{56}$ |
| −20.0 | $8.4 \times 10^{57}$ | $1.0 \times 10^{58}$ | $1.5 \times 10^{58}$ | $1.7 \times 10^{58}$ | $1.1 \times 10^{58}$ | $5.4 \times 10^{57}$ | $2.4 \times 10^{57}$ | $1.1 \times 10^{57}$ | $4.6 \times 10^{56}$ | $2.0 \times 10^{56}$ |

(Table continued)

(Continued)

| $\log_{10} y$ | $m_X = 5$ MeV | 6 MeV | 8 MeV | 9 MeV | 11 MeV | 14 MeV | 17 MeV | 21 MeV | 26 MeV | 32 MeV |
|---|---|---|---|---|---|---|---|---|---|---|
| −20.3 | $1.1 \times 10^{58}$ | $1.5 \times 10^{58}$ | $1.8 \times 10^{58}$ | $1.2 \times 10^{58}$ | $5.6 \times 10^{57}$ | $2.5 \times 10^{57}$ | $1.1 \times 10^{57}$ | $4.9 \times 10^{56}$ | $2.1 \times 10^{56}$ | $9.4 \times 10^{55}$ |
| −20.7 | $1.5 \times 10^{58}$ | $1.8 \times 10^{58}$ | $1.3 \times 10^{58}$ | $5.9 \times 10^{57}$ | $2.6 \times 10^{57}$ | $1.2 \times 10^{57}$ | $5.1 \times 10^{56}$ | $2.3 \times 10^{56}$ | $1.0 \times 10^{56}$ | $4.4 \times 10^{55}$ |
| −21.0 | $1.8 \times 10^{58}$ | $1.3 \times 10^{58}$ | $6.2 \times 10^{57}$ | $2.7 \times 10^{57}$ | $1.2 \times 10^{57}$ | $5.4 \times 10^{56}$ | $2.4 \times 10^{56}$ | $1.1 \times 10^{56}$ | $4.6 \times 10^{55}$ | $2.0 \times 10^{55}$ |
| −21.3 | $1.4 \times 10^{58}$ | $6.5 \times 10^{57}$ | $2.9 \times 10^{57}$ | $1.3 \times 10^{57}$ | $5.6 \times 10^{56}$ | $2.5 \times 10^{56}$ | $1.1 \times 10^{56}$ | $4.9 \times 10^{55}$ | $2.1 \times 10^{55}$ | $9.4 \times 10^{54}$ |
| −21.7 | $6.8 \times 10^{57}$ | $3.0 \times 10^{57}$ | $1.3 \times 10^{57}$ | $5.9 \times 10^{56}$ | $2.6 \times 10^{56}$ | $1.2 \times 10^{56}$ | $5.1 \times 10^{55}$ | $2.3 \times 10^{55}$ | $1.0 \times 10^{55}$ | $4.4 \times 10^{54}$ |
| −22.0 | $3.2 \times 10^{57}$ | $1.4 \times 10^{57}$ | $6.2 \times 10^{56}$ | $2.7 \times 10^{56}$ | $1.2 \times 10^{56}$ | $5.4 \times 10^{55}$ | $2.4 \times 10^{55}$ | $1.1 \times 10^{55}$ | $4.6 \times 10^{54}$ | $2.0 \times 10^{54}$ |

| $\log_{10} y$ | $m_X = 39$ MeV | 48 MeV | 58 MeV | 88 MeV | 108 MeV | 132 MeV | 162 MeV | 199 MeV | 244 MeV | 300 MeV |
|---|---|---|---|---|---|---|---|---|---|---|
| −13.3 | $1.2 \times 10^{53}$ | $2.7 \times 10^{53}$ | $2.0 \times 10^{53}$ | $5.1 \times 10^{53}$ | $2.1 \times 10^{53}$ | $2.3 \times 10^{53}$ | $3.0 \times 10^{53}$ | $2.0 \times 10^{53}$ | $7.1 \times 10^{52}$ | $5.1 \times 10^{52}$ |
| −13.7 | $1.6 \times 10^{54}$ | $1.0 \times 10^{54}$ | $1.4 \times 10^{54}$ | $1.3 \times 10^{54}$ | $1.0 \times 10^{54}$ | $9.9 \times 10^{53}$ | $9.5 \times 10^{53}$ | $5.6 \times 10^{53}$ | $2.7 \times 10^{53}$ | $3.4 \times 10^{53}$ |
| −14.0 | $6.6 \times 10^{54}$ | $4.1 \times 10^{54}$ | $7.8 \times 10^{54}$ | $3.7 \times 10^{54}$ | $4.8 \times 10^{54}$ | $4.7 \times 10^{54}$ | $3.2 \times 10^{54}$ | $1.3 \times 10^{54}$ | $1.0 \times 10^{54}$ | $1.8 \times 10^{54}$ |
| −14.3 | $1.7 \times 10^{55}$ | $3.9 \times 10^{55}$ | $2.3 \times 10^{55}$ | $1.8 \times 10^{55}$ | $1.9 \times 10^{55}$ | $1.4 \times 10^{55}$ | $6.9 \times 10^{54}$ | $5.3 \times 10^{54}$ | $6.7 \times 10^{54}$ | $4.4 \times 10^{54}$ |
| −14.7 | $5.8 \times 10^{55}$ | $3.9 \times 10^{55}$ | $4.6 \times 10^{55}$ | $6.2 \times 10^{55}$ | $5.3 \times 10^{55}$ | $3.4 \times 10^{55}$ | $1.8 \times 10^{55}$ | $1.9 \times 10^{55}$ | $2.3 \times 10^{55}$ | $4.6 \times 10^{54}$ |
| −15.0 | $1.4 \times 10^{56}$ | $2.1 \times 10^{56}$ | $1.4 \times 10^{56}$ | $1.6 \times 10^{56}$ | $1.3 \times 10^{56}$ | $8.6 \times 10^{55}$ | $7.5 \times 10^{55}$ | $8.7 \times 10^{55}$ | $3.3 \times 10^{55}$ | $2.5 \times 10^{54}$ |
| −15.3 | $4.6 \times 10^{56}$ | $3.7 \times 10^{56}$ | $4.7 \times 10^{56}$ | $3.9 \times 10^{56}$ | $2.7 \times 10^{56}$ | $2.2 \times 10^{56}$ | $2.7 \times 10^{56}$ | $1.5 \times 10^{56}$ | $2.2 \times 10^{55}$ | $1.2 \times 10^{54}$ |
| −15.7 | $8.4 \times 10^{56}$ | $9.5 \times 10^{56}$ | $1.0 \times 10^{57}$ | $6.9 \times 10^{56}$ | $6.5 \times 10^{56}$ | $6.6 \times 10^{56}$ | $4.7 \times 10^{56}$ | $1.1 \times 10^{56}$ | $1.0 \times 10^{55}$ | $5.4 \times 10^{53}$ |
| −16.0 | $1.6 \times 10^{57}$ | $1.6 \times 10^{57}$ | $1.7 \times 10^{57}$ | $1.6 \times 10^{57}$ | $1.5 \times 10^{57}$ | $1.2 \times 10^{57}$ | $3.7 \times 10^{56}$ | $5.4 \times 10^{55}$ | $4.8 \times 10^{54}$ | $2.5 \times 10^{53}$ |
| −16.3 | $2.4 \times 10^{57}$ | $2.7 \times 10^{57}$ | $2.8 \times 10^{57}$ | $3.0 \times 10^{57}$ | $2.3 \times 10^{57}$ | $8.5 \times 10^{56}$ | $1.8 \times 10^{56}$ | $2.5 \times 10^{55}$ | $2.2 \times 10^{54}$ | $1.2 \times 10^{53}$ |
| −16.7 | $3.6 \times 10^{57}$ | $4.1 \times 10^{57}$ | $4.6 \times 10^{57}$ | $3.7 \times 10^{57}$ | $1.5 \times 10^{57}$ | $4.2 \times 10^{56}$ | $8.4 \times 10^{55}$ | $1.2 \times 10^{55}$ | $1.0 \times 10^{54}$ | $5.4 \times 10^{52}$ |
| −17.0 | $5.3 \times 10^{57}$ | $6.1 \times 10^{57}$ | $6.8 \times 10^{57}$ | $2.3 \times 10^{57}$ | $7.5 \times 10^{56}$ | $2.0 \times 10^{56}$ | $3.9 \times 10^{55}$ | $5.4 \times 10^{54}$ | $4.8 \times 10^{53}$ | $2.5 \times 10^{52}$ |
| −17.3 | $8.0 \times 10^{57}$ | $8.8 \times 10^{57}$ | $6.4 \times 10^{57}$ | $1.1 \times 10^{57}$ | $3.5 \times 10^{56}$ | $9.1 \times 10^{55}$ | $1.8 \times 10^{55}$ | $2.5 \times 10^{54}$ | $2.2 \times 10^{53}$ | $1.2 \times 10^{52}$ |
| −17.7 | $1.1 \times 10^{58}$ | $7.4 \times 10^{57}$ | $3.4 \times 10^{57}$ | $5.1 \times 10^{56}$ | $1.6 \times 10^{56}$ | $4.2 \times 10^{55}$ | $8.4 \times 10^{54}$ | $1.2 \times 10^{54}$ | $1.0 \times 10^{53}$ | $5.4 \times 10^{51}$ |
| −18.0 | $8.2 \times 10^{57}$ | $3.8 \times 10^{57}$ | $1.6 \times 10^{57}$ | $2.4 \times 10^{56}$ | $7.5 \times 10^{55}$ | $2.0 \times 10^{55}$ | $3.9 \times 10^{54}$ | $5.4 \times 10^{53}$ | $4.8 \times 10^{52}$ | $2.5 \times 10^{51}$ |
| −18.3 | $4.1 \times 10^{57}$ | $1.8 \times 10^{57}$ | $7.4 \times 10^{56}$ | $1.1 \times 10^{56}$ | $3.5 \times 10^{55}$ | $9.1 \times 10^{54}$ | $1.8 \times 10^{54}$ | $2.5 \times 10^{53}$ | $2.2 \times 10^{52}$ | $1.2 \times 10^{51}$ |
| −18.7 | $1.9 \times 10^{57}$ | $8.2 \times 10^{56}$ | $3.4 \times 10^{56}$ | $5.1 \times 10^{55}$ | $1.6 \times 10^{55}$ | $4.2 \times 10^{54}$ | $8.4 \times 10^{53}$ | $1.2 \times 10^{53}$ | $1.0 \times 10^{52}$ | $5.4 \times 10^{50}$ |
| −19.0 | $8.9 \times 10^{56}$ | $3.8 \times 10^{56}$ | $1.6 \times 10^{56}$ | $2.4 \times 10^{55}$ | $7.5 \times 10^{54}$ | $2.0 \times 10^{54}$ | $3.9 \times 10^{53}$ | $5.4 \times 10^{52}$ | $4.8 \times 10^{51}$ | $2.5 \times 10^{50}$ |
| −19.3 | $4.1 \times 10^{56}$ | $1.8 \times 10^{56}$ | $7.4 \times 10^{55}$ | $1.1 \times 10^{55}$ | $3.5 \times 10^{54}$ | $9.1 \times 10^{53}$ | $1.8 \times 10^{53}$ | $2.5 \times 10^{52}$ | $2.2 \times 10^{51}$ | $1.2 \times 10^{50}$ |
| −19.7 | $1.9 \times 10^{56}$ | $8.2 \times 10^{55}$ | $3.4 \times 10^{55}$ | $5.1 \times 10^{54}$ | $1.6 \times 10^{54}$ | $4.2 \times 10^{53}$ | $8.4 \times 10^{52}$ | $1.2 \times 10^{52}$ | $1.0 \times 10^{51}$ | $5.4 \times 10^{49}$ |
| −20.0 | $8.9 \times 10^{55}$ | $3.8 \times 10^{55}$ | $1.6 \times 10^{55}$ | $2.4 \times 10^{54}$ | $7.5 \times 10^{53}$ | $2.0 \times 10^{53}$ | $3.9 \times 10^{52}$ | $5.4 \times 10^{51}$ | $4.8 \times 10^{50}$ | $2.5 \times 10^{49}$ |
| −20.3 | $4.1 \times 10^{55}$ | $1.8 \times 10^{55}$ | $7.4 \times 10^{54}$ | $1.1 \times 10^{54}$ | $3.5 \times 10^{53}$ | $9.1 \times 10^{52}$ | $1.8 \times 10^{52}$ | $2.5 \times 10^{51}$ | $2.2 \times 10^{50}$ | $1.2 \times 10^{49}$ |
| −20.7 | $1.9 \times 10^{55}$ | $8.2 \times 10^{54}$ | $3.4 \times 10^{54}$ | $5.1 \times 10^{53}$ | $1.6 \times 10^{53}$ | $4.2 \times 10^{52}$ | $8.4 \times 10^{51}$ | $1.2 \times 10^{51}$ | $1.0 \times 10^{50}$ | $5.4 \times 10^{48}$ |
| −21.0 | $8.9 \times 10^{54}$ | $3.8 \times 10^{54}$ | $1.6 \times 10^{54}$ | $2.4 \times 10^{53}$ | $7.5 \times 10^{52}$ | $2.0 \times 10^{52}$ | $3.9 \times 10^{51}$ | $5.4 \times 10^{50}$ | $4.8 \times 10^{49}$ | $2.5 \times 10^{48}$ |
| −21.3 | $4.1 \times 10^{54}$ | $1.8 \times 10^{54}$ | $7.4 \times 10^{53}$ | $1.1 \times 10^{53}$ | $3.5 \times 10^{52}$ | $9.1 \times 10^{51}$ | $1.8 \times 10^{51}$ | $2.5 \times 10^{50}$ | $2.2 \times 10^{49}$ | $1.2 \times 10^{48}$ |
| −21.7 | $1.9 \times 10^{54}$ | $8.2 \times 10^{53}$ | $3.4 \times 10^{53}$ | $5.1 \times 10^{52}$ | $1.6 \times 10^{52}$ | $4.2 \times 10^{51}$ | $8.4 \times 10^{50}$ | $1.2 \times 10^{50}$ | $1.0 \times 10^{49}$ | $5.4 \times 10^{47}$ |
| −22.0 | $8.9 \times 10^{53}$ | $3.8 \times 10^{53}$ | $1.6 \times 10^{53}$ | $2.4 \times 10^{52}$ | $7.5 \times 10^{51}$ | $2.0 \times 10^{51}$ | $3.9 \times 10^{50}$ | $5.4 \times 10^{49}$ | $4.8 \times 10^{48}$ | $2.5 \times 10^{47}$ |

### 2. Annihilation sphere

These tables contain entries for $r_N$, the radius (in km) at which the number-changing processes cease. "Bulk" indicates that the diffusive approximation breaks down and there is no defined annihilation sphere.

| $\log_{10} y$ | $m_X = 5$ MeV | 6 MeV | 8 MeV | 9 MeV | 11 MeV | 14 MeV | 17 MeV | 21 MeV | 26 MeV | 32 MeV |
|---|---|---|---|---|---|---|---|---|---|---|
| −13.3 | 174.0 | 142.0 | 116.0 | 94.4 | 77.0 | 62.6 | 50.9 | 41.2 | 33.4 | 27.1 |
| −13.7 | 153.0 | 125.0 | 102.0 | 82.8 | 67.4 | 54.8 | 44.4 | 36.0 | 29.1 | 23.8 |
| −14.0 | 133.0 | 108.0 | 88.4 | 71.9 | 58.4 | 47.5 | 38.4 | 31.0 | 25.3 | 21.3 |
| −14.3 | 114.0 | 92.7 | 75.5 | 61.5 | 49.9 | 40.5 | 32.7 | 26.5 | 22.1 | 20.5 |
| −14.7 | 96.6 | 78.7 | 63.9 | 51.9 | 42.1 | 34.0 | 27.6 | 22.8 | 20.6 | 19.9 |
| −15.0 | 80.6 | 65.6 | 53.3 | 43.2 | 35.0 | 28.3 | 23.3 | 20.7 | 20.0 | 19.2 |
| −15.3 | 66.6 | 54.2 | 43.8 | 35.5 | 28.7 | 23.6 | 20.8 | 20.1 | 19.3 | 18.6 |
| −15.7 | 54.2 | 44.1 | 35.6 | 28.7 | 23.6 | 20.8 | 20.1 | 19.4 | 18.7 | 18.0 |

(Table continued)

(Continued)

| $\log_{10} y$ | $m_X = 5$ MeV | 6 MeV | 8 MeV | 9 MeV | 11 MeV | 14 MeV | 17 MeV | 21 MeV | 26 MeV | 32 MeV |
|---|---|---|---|---|---|---|---|---|---|---|
| −16.0 | 43.5 | 35.1 | 28.5 | 23.4 | 20.8 | 20.1 | 19.4 | 18.7 | 18.0 | 17.4 |
| −16.3 | 34.3 | 27.8 | 22.9 | 20.7 | 20.0 | 19.3 | 18.6 | 18.0 | 17.3 | 16.7 |
| −16.7 | 26.8 | 22.3 | 20.6 | 19.9 | 19.2 | 18.6 | 17.9 | 17.3 | 16.6 | 16.0 |
| −17.0 | 21.6 | 20.4 | 19.7 | 19.1 | 18.4 | 17.8 | 17.2 | 16.5 | 15.9 | 15.3 |
| −17.3 | 20.3 | 19.6 | 18.9 | 18.3 | 17.6 | 17.0 | 16.4 | 15.8 | 15.2 | 13.9 |
| −17.7 | 19.4 | 18.8 | 18.1 | 17.5 | 16.9 | 16.3 | 15.7 | 15.0 | 13.6 | 11.8 |
| −18.0 | 18.6 | 18.0 | 17.3 | 16.7 | 16.1 | 15.5 | 14.7 | 13.1 | 11.4 | Bulk |
| −18.3 | 17.8 | 17.2 | 16.5 | 15.9 | 15.3 | 14.3 | 12.6 | 10.9 | Bulk | Bulk |
| −18.7 | 17.0 | 16.4 | 15.8 | 15.2 | 13.8 | 12.1 | 10.4 | Bulk | Bulk | Bulk |
| −19.0 | 16.2 | 15.6 | 15.0 | 13.4 | 11.7 | Bulk | Bulk | Bulk | Bulk | Bulk |
| −19.3 | 15.4 | 14.5 | 12.9 | 11.1 | Bulk | Bulk | Bulk | Bulk | Bulk | Bulk |
| −19.7 | 14.1 | 12.4 | 10.6 | Bulk | Bulk | Bulk | Bulk | Bulk | Bulk | Bulk |
| −20.0 | 11.9 | 10.1 | Bulk | Bulk | Bulk | Bulk | Bulk | Bulk | Bulk | Bulk |
| −20.3 | Bulk | Bulk | Bulk | Bulk | Bulk | Bulk | Bulk | Bulk | Bulk | Bulk |
| −20.7 | Bulk | Bulk | Bulk | Bulk | Bulk | Bulk | Bulk | Bulk | Bulk | Bulk |
| −21.0 | Bulk | Bulk | Bulk | Bulk | Bulk | Bulk | Bulk | Bulk | Bulk | Bulk |
| −21.3 | Bulk | Bulk | Bulk | Bulk | Bulk | Bulk | Bulk | Bulk | Bulk | Bulk |
| −21.7 | Bulk | Bulk | Bulk | Bulk | Bulk | Bulk | Bulk | Bulk | Bulk | Bulk |
| −22.0 | Bulk | Bulk | Bulk | Bulk | Bulk | Bulk | Bulk | Bulk | Bulk | Bulk |

| $\log_{10} y$ | $m_X = 39$ MeV | 48 MeV | 58 MeV | 88 MeV | 108 MeV | 132 MeV | 162 MeV | 199 MeV | 244 MeV | 300 MeV |
|---|---|---|---|---|---|---|---|---|---|---|
| −13.3 | 20.5 | 19.7 | 18.9 | 18.2 | 17.5 | 16.7 | 16.0 | 15.3 | 14.0 | 12.0 |
| −13.7 | 20.0 | 19.2 | 18.5 | 17.8 | 17.0 | 16.3 | 15.6 | 14.8 | 13.0 | 11.0 |
| −14.0 | 19.5 | 18.7 | 18.0 | 17.3 | 16.6 | 15.9 | 15.2 | 13.8 | 11.9 | Bulk |
| −14.3 | 18.9 | 18.2 | 17.5 | 16.8 | 16.1 | 15.4 | 14.5 | 12.7 | 10.7 | Bulk |
| −14.7 | 18.4 | 17.7 | 17.0 | 16.3 | 15.7 | 15.0 | 13.3 | 11.4 | Bulk | Bulk |
| −15.0 | 17.8 | 17.2 | 16.5 | 15.8 | 15.2 | 13.8 | 12.0 | 10.1 | Bulk | Bulk |
| −15.3 | 17.3 | 16.6 | 15.9 | 15.3 | 14.1 | 12.4 | 10.5 | Bulk | Bulk | Bulk |
| −15.7 | 16.7 | 16.0 | 15.4 | 14.4 | 12.7 | 10.8 | Bulk | Bulk | Bulk | Bulk |
| −16.0 | 16.1 | 15.4 | 14.5 | 12.8 | 11.0 | Bulk | Bulk | Bulk | Bulk | Bulk |
| −16.3 | 15.4 | 14.5 | 12.9 | 11.1 | Bulk | Bulk | Bulk | Bulk | Bulk | Bulk |
| −16.7 | 14.4 | 12.8 | 11.0 | Bulk | Bulk | Bulk | Bulk | Bulk | Bulk | Bulk |
| −17.0 | 12.6 | 10.8 | Bulk | Bulk | Bulk | Bulk | Bulk | Bulk | Bulk | Bulk |
| −17.3 | 10.5 | Bulk | Bulk | Bulk | Bulk | Bulk | Bulk | Bulk | Bulk | Bulk |
| −17.7 | Bulk | Bulk | Bulk | Bulk | Bulk | Bulk | Bulk | Bulk | Bulk | Bulk |
| −18.0 | Bulk | Bulk | Bulk | Bulk | Bulk | Bulk | Bulk | Bulk | Bulk | Bulk |
| −18.3 | Bulk | Bulk | Bulk | Bulk | Bulk | Bulk | Bulk | Bulk | Bulk | Bulk |
| −18.7 | Bulk | Bulk | Bulk | Bulk | Bulk | Bulk | Bulk | Bulk | Bulk | Bulk |
| −19.0 | Bulk | Bulk | Bulk | Bulk | Bulk | Bulk | Bulk | Bulk | Bulk | Bulk |
| −19.3 | Bulk | Bulk | Bulk | Bulk | Bulk | Bulk | Bulk | Bulk | Bulk | Bulk |
| −19.7 | Bulk | Bulk | Bulk | Bulk | Bulk | Bulk | Bulk | Bulk | Bulk | Bulk |
| −20.0 | Bulk | Bulk | Bulk | Bulk | Bulk | Bulk | Bulk | Bulk | Bulk | Bulk |
| −20.3 | Bulk | Bulk | Bulk | Bulk | Bulk | Bulk | Bulk | Bulk | Bulk | Bulk |
| −20.7 | Bulk | Bulk | Bulk | Bulk | Bulk | Bulk | Bulk | Bulk | Bulk | Bulk |
| −21.0 | Bulk | Bulk | Bulk | Bulk | Bulk | Bulk | Bulk | Bulk | Bulk | Bulk |
| −21.3 | Bulk | Bulk | Bulk | Bulk | Bulk | Bulk | Bulk | Bulk | Bulk | Bulk |
| −21.7 | Bulk | Bulk | Bulk | Bulk | Bulk | Bulk | Bulk | Bulk | Bulk | Bulk |
| −22.0 | Bulk | Bulk | Bulk | Bulk | Bulk | Bulk | Bulk | Bulk | Bulk | Bulk |

### 3. Energy sphere

These tables contain entries for $r_E$, the radius (in km) at which the processes that keep the SM and DM in thermal contact cease. "Bulk" indicates that the diffusive approximation breaks down and there is no defined energy sphere.

| $\log_{10} y$ | $m_X = 5$ MeV | 6 MeV | 8 MeV | 9 MeV | 11 MeV | 14 MeV | 17 MeV | 21 MeV | 26 MeV | 32 MeV |
|---|---|---|---|---|---|---|---|---|---|---|
| −13.3 | 300.0 | 248.0 | 203.0 | 167.0 | 136.0 | 111.0 | 89.8 | 72.8 | 59.3 | 48.2 |
| −13.7 | 233.0 | 192.0 | 156.0 | 127.0 | 104.0 | 84.4 | 69.0 | 55.9 | 45.5 | 37.1 |
| −14.0 | 181.0 | 148.0 | 121.0 | 98.5 | 80.1 | 65.1 | 53.0 | 43.1 | 35.1 | 28.6 |
| −14.3 | 141.0 | 114.0 | 93.6 | 76.2 | 62.2 | 50.6 | 41.2 | 33.6 | 27.2 | 22.9 |
| −14.7 | 110.0 | 89.3 | 72.8 | 59.5 | 48.4 | 39.5 | 32.1 | 26.1 | 22.4 | 20.8 |
| −15.0 | 86.1 | 70.4 | 57.4 | 46.7 | 38.0 | 31.0 | 25.1 | 22.1 | 20.6 | 19.7 |
| −15.3 | 68.0 | 55.7 | 45.3 | 36.9 | 30.1 | 24.4 | 21.8 | 20.4 | 19.6 | 18.9 |
| −15.7 | 54.2 | 44.2 | 36.1 | 29.3 | 23.9 | 21.6 | 20.3 | 19.5 | 18.9 | 18.3 |
| −16.0 | 43.4 | 35.4 | 28.9 | 23.6 | 21.4 | 20.2 | 19.4 | 18.8 | 18.2 | 17.7 |
| −16.3 | 34.9 | 28.4 | 23.4 | 21.3 | 20.1 | 19.3 | 18.7 | 18.1 | 17.6 | 17.1 |
| −16.7 | 28.1 | 23.2 | 21.1 | 20.0 | 19.3 | 18.7 | 18.1 | 17.6 | 17.0 | 16.5 |
| −17.0 | 23.2 | 21.1 | 20.0 | 19.2 | 18.6 | 18.1 | 17.5 | 17.0 | 16.5 | 15.9 |
| −17.3 | 21.0 | 19.9 | 19.2 | 18.6 | 18.0 | 17.5 | 16.9 | 16.4 | 15.9 | 15.4 |
| −17.7 | 19.9 | 19.2 | 18.6 | 18.0 | 17.5 | 16.9 | 16.4 | 15.9 | 15.3 | 14.8 |
| −18.0 | 19.2 | 18.6 | 18.0 | 17.4 | 16.9 | 16.4 | 15.8 | 15.3 | 14.8 | 13.9 |
| −18.3 | 18.6 | 18.0 | 17.4 | 16.9 | 16.4 | 15.8 | 15.3 | 14.7 | 13.8 | 12.4 |
| −18.7 | 18.0 | 17.4 | 16.9 | 16.4 | 15.8 | 15.3 | 14.7 | 13.8 | 12.4 | 10.4 |
| −19.0 | 17.4 | 16.9 | 16.4 | 15.8 | 15.3 | 14.7 | 13.8 | 12.3 | 10.4 | Bulk |
| −19.3 | 16.9 | 16.4 | 15.8 | 15.3 | 14.7 | 13.8 | 12.3 | 10.3 | Bulk | Bulk |
| −19.7 | 16.4 | 15.8 | 15.3 | 14.7 | 13.8 | 12.3 | 10.4 | Bulk | Bulk | Bulk |
| −20.0 | 15.9 | 15.3 | 14.8 | 13.8 | 12.4 | 10.4 | Bulk | Bulk | Bulk | Bulk |
| −20.3 | 15.3 | 14.8 | 13.9 | 12.4 | 10.4 | Bulk | Bulk | Bulk | Bulk | Bulk |
| −20.7 | 14.8 | 13.9 | 12.4 | 10.5 | Bulk | Bulk | Bulk | Bulk | Bulk | Bulk |
| −21.0 | 14.0 | 12.5 | 10.6 | Bulk | Bulk | Bulk | Bulk | Bulk | Bulk | Bulk |
| −21.3 | 12.6 | 10.7 | Bulk | Bulk | Bulk | Bulk | Bulk | Bulk | Bulk | Bulk |
| −21.7 | 10.8 | Bulk | Bulk | Bulk | Bulk | Bulk | Bulk | Bulk | Bulk | Bulk |
| −22.0 | Bulk | Bulk | Bulk | Bulk | Bulk | Bulk | Bulk | Bulk | Bulk | Bulk |

| $\log_{10} y$ | $m_X = 39$ MeV | 48 MeV | 58 MeV | 88 MeV | 108 MeV | 132 MeV | 162 MeV | 199 MeV | 244 MeV | 300 MeV |
|---|---|---|---|---|---|---|---|---|---|---|
| −13.3 | 31.9 | 25.7 | 22.4 | 20.8 | 19.8 | 19.2 | 18.6 | 18.0 | 17.5 | 16.9 |
| −13.7 | 24.3 | 21.9 | 20.5 | 19.7 | 19.0 | 18.4 | 17.9 | 17.3 | 16.8 | 16.3 |
| −14.0 | 21.5 | 20.3 | 19.5 | 18.8 | 18.3 | 17.7 | 17.2 | 16.7 | 16.2 | 15.7 |
| −14.3 | 20.0 | 19.3 | 18.7 | 18.1 | 17.6 | 17.1 | 16.6 | 16.1 | 15.5 | 15.0 |
| −14.7 | 19.2 | 18.6 | 18.0 | 17.5 | 17.0 | 16.5 | 16.0 | 15.4 | 14.9 | 14.2 |
| −15.0 | 18.5 | 17.9 | 17.4 | 16.9 | 16.4 | 15.9 | 15.3 | 14.8 | 14.0 | 12.8 |
| −15.3 | 17.8 | 17.3 | 16.8 | 16.3 | 15.8 | 15.2 | 14.7 | 13.8 | 12.5 | 10.7 |
| −15.7 | 17.2 | 16.7 | 16.2 | 15.7 | 15.2 | 14.6 | 13.7 | 12.2 | 10.3 | Bulk |
| −16.0 | 16.6 | 16.1 | 15.6 | 15.1 | 14.5 | 13.4 | 11.9 | Bulk | Bulk | Bulk |
| −16.3 | 16.0 | 15.5 | 15.0 | 14.4 | 13.3 | 11.7 | Bulk | Bulk | Bulk | Bulk |
| −16.7 | 15.5 | 14.9 | 14.2 | 13.1 | 11.4 | Bulk | Bulk | Bulk | Bulk | Bulk |
| −17.0 | 14.9 | 14.1 | 12.9 | 11.1 | Bulk | Bulk | Bulk | Bulk | Bulk | Bulk |
| −17.3 | 14.0 | 12.7 | 11.0 | Bulk | Bulk | Bulk | Bulk | Bulk | Bulk | Bulk |
| −17.7 | 12.6 | 10.8 | Bulk | Bulk | Bulk | Bulk | Bulk | Bulk | Bulk | Bulk |
| −18.0 | 10.6 | Bulk | Bulk | Bulk | Bulk | Bulk | Bulk | Bulk | Bulk | Bulk |
| −18.3 | Bulk | Bulk | Bulk | Bulk | Bulk | Bulk | Bulk | Bulk | Bulk | Bulk |
| −18.7 | Bulk | Bulk | Bulk | Bulk | Bulk | Bulk | Bulk | Bulk | Bulk | Bulk |
| −19.0 | Bulk | Bulk | Bulk | Bulk | Bulk | Bulk | Bulk | Bulk | Bulk | Bulk |
| −19.3 | Bulk | Bulk | Bulk | Bulk | Bulk | Bulk | Bulk | Bulk | Bulk | Bulk |
| −19.7 | Bulk | Bulk | Bulk | Bulk | Bulk | Bulk | Bulk | Bulk | Bulk | Bulk |
| −20.0 | Bulk | Bulk | Bulk | Bulk | Bulk | Bulk | Bulk | Bulk | Bulk | Bulk |

(Continued)

| $\log_{10} y$ | $m_X =$ 39 MeV | 48 MeV | 58 MeV | 88 MeV | 108 MeV | 132 MeV | 162 MeV | 199 MeV | 244 MeV | 300 MeV |
|---|---|---|---|---|---|---|---|---|---|---|
| −20.3 | Bulk | Bulk | Bulk | Bulk | Bulk | Bulk | Bulk | Bulk | Bulk | Bulk |
| −20.7 | Bulk | Bulk | Bulk | Bulk | Bulk | Bulk | Bulk | Bulk | Bulk | Bulk |
| −21.0 | Bulk | Bulk | Bulk | Bulk | Bulk | Bulk | Bulk | Bulk | Bulk | Bulk |
| −21.3 | Bulk | Bulk | Bulk | Bulk | Bulk | Bulk | Bulk | Bulk | Bulk | Bulk |
| −21.7 | Bulk | Bulk | Bulk | Bulk | Bulk | Bulk | Bulk | Bulk | Bulk | Bulk |
| −22.0 | Bulk | Bulk | Bulk | Bulk | Bulk | Bulk | Bulk | Bulk | Bulk | Bulk |